%% file: main.tex
\PassOptionsToPackage{bookmarks={false}}{hyperref}
\documentclass[conference]{IEEEtran}
\usepackage{amsmath}
\usepackage{mathtools}
\usepackage{color}
\usepackage{cite}
\usepackage{hyperref}
\usepackage[english]{babel}
\usepackage{amssymb}
\usepackage{pifont}
\usepackage{textcomp}
\usepackage{tikz}
\usepackage{graphicx}
\usepackage{caption}
\usepackage{subcaption}
\usepackage{arydshln}
\usepackage{tikz}
\usepackage{mathrsfs}
\usepackage{wasysym}
\usepackage{algorithm}
\usepackage{algorithmicx}
\usepackage{algpseudocode}
\usepackage{comment}
\usepackage[utf8]{inputenc}
\usepackage{multirow}
\usepackage{cleveref}

\algdef{SE}[DOWHILE]{Do}{doWhile}{\algorithmicdo}[1]{\algorithmicwhile\ #1}%
\algrenewcommand\textproc{}

\newcommand{\xmark}{\ding{55}}
\newcommand{\fig}[1]{Fig.~\ref{fig:#1}}
\newcommand{\tabl}[1]{Table~\ref{table:#1}}
\newcommand{\alg}[1]{Alg.~\ref{algorithm:#1}}
\newcommand{\sect}[1]{Section~\ref{section:#1}}
\newcommand{\eqn}[1]{Eq.~\ref{equation:#1}}

\begin{document}
\setlength{\abovedisplayskip}{2pt}
\setlength{\belowdisplayskip}{2pt}
\setlength\floatsep{1.25\baselineskip plus 3pt minus 2pt}
\setlength\textfloatsep{1.25\baselineskip plus 3pt minus 2pt}
\setlength\intextsep{1.25\baselineskip plus 3pt minus 2 pt}

\input{sections/title}

\input{sections/abstract}

\input{sections/introduction}

\input{sections/relatedWorks.tex}

\input{sections/challenges}

\input{sections/problem}

\input{sections/solution}

\input{sections/evaluation}

\input{sections/conclusion}


\bibliographystyle{plain}
\bibliography{library}


\end{document}

%% file: sections/title.tex
\title{Service Function Chaining Simplified\vspace{-0.5cm}}
\author{\IEEEauthorblockN{Milad Ghaznavi, Nashid Shahriar, Reaz Ahmed, Raouf Boutaba}
\IEEEauthorblockA{David R. Cheriton School of Computer Science, University of Waterloo, ON, Canada\\
\texttt{\{eghaznav | nshahria | r5ahmed | rboutaba\}@uwaterloo.ca}}\vspace{-1.0cm}}
\maketitle

%% file: sections/abstract.tex
\begin{abstract}\label{section:abstract}
Middleboxes have become a vital part of modern networks by providing \textit{service functions} such as content filtering, load balancing and optimization of network traffic. An ordered sequence of middleboxes composing a logical service is called \textit{service chain}. \textit{Service Function Chaining} (SFC) enables us to define these service chains. Recent optimization models of SFCs assume that the functionality of a middlebox is provided by a single software appliance, commonly known as \textit{Virtual Network Function} (VNF). This assumption limits SFCs to the throughput of an individual VNF and resources of a physical machine hosting the \textit{VNF instance}. Moreover, typical service providers offer VNFs with heterogeneous throughput and resource configurations. Thus, deploying a service chain with custom throughput can become a tedious process of stitching heterogeneous VNF instances. In this paper, we describe how we can overcome these limitations without worrying about underlying VNF configurations and resource constraints. This prospect is achieved by distributed deploying multiple VNF instances providing the functionality of a middlebox and modeling the optimal deployment of a service chain as a mixed integer programming problem. The proposed model optimizes host and bandwidth resources allocation, and determines the optimal placement of VNF instances, while balancing workload and routing traffic among these VNF instances. We show that this problem is NP-Hard and propose a heuristic solution
called \textit{Kariz}. Kariz utilizes a tuning parameter to control the trade-off between speed and accuracy of the solution.
Finally, our solution is evaluated using simulations in data-center networks.
\end{abstract}

%% file: sections/introduction.tex
\section{Introduction}\label{section:introduction}
Network Function Virtualization (NFV) is expected to instigate a revolutionary change in the networking industry. This industry still has the “mainframe” mindset relying on vendor specific, proprietary middleboxes providing various network functions. Examples of such middleboxes include firewalls, proxies, WAN optimizers, Intrusion Detection Systems (IDSs), etc. NFV proposes to replace proprietary, hardware middleboxes with innovative and flexible software middleboxes also known as Virtual Network Functions (VNFs).

VNFs are generally run on commodity (e.g., x86 based systems) hardware. In this way, the capital and operational expenditures of buying and maintaining specialized hardware is reduced. However, VNFs are yet to achieve the same performance of their hardware counterparts. This impedes the real life adoption of VNFs in today's networks carrying voluminous data traffic every second. In these networks, traffic is often required to pass through and processed by an ordered sequence of VNFs called \textit{service chain}. For instance, traffic may need to pass through an IDS, then a proxy, and finally through a firewall. This phenomenon is commonly referred to as \textit{Service Function Chaining} (SFC) \cite{rfc-sfc}. Service chains or simply chains are required to process large volumes of traffic within a very short period of time to facilitate real-time streaming applications that comprise majority of traffic in today's networks. Failure to provide the desired throughput of a chain may lead to violation of the service level agreements incurring high penalties. Hence, achieving high throughput of VNFs is of paramount importance.

There are several streams of on going researches towards increasing the throughput of a VNF. The first stream explores the possibility to build virtual platforms capable of processing packets very fast by utilizing advanced hardware technologies \cite{hwang2015netvm}, \cite{martins2014clickos}. The second stream combines VNFs with hardware middleboxes to facilitate a better usability of the existing hardware middleboxes \cite{moens2014vnf} and brings the benefits of the both worlds. However, none of these approaches can overcome the physical limitation of {deploying a VNF on a single physical machine}. Hence, the third stream of works including \cite{opennf}, \cite{splitMerge} propose to redistribute the traffic destined to a VNF across multiple VNF instances running independently on {different CPU cores of a server, or even different servers} altogether providing the functionality of the VNF. The cluster architecture of Bro IDS \cite{broIDS} is an example of such distributed deployment. In addition to achieving higher throughput, the distributed deployment offers better flexibility and reliability of the deployed chains than the standalone counterpart.

A fundamental problem for deploying a chain with a custom throughput is the resource efficient selection and placement of VNF instances. Solving this problem requires addressing several optimization challenges. First, there can be heterogeneity in terms of the throughput of different VNF instances. For instance, virtual WAN optimizer such as Riverbed STEELHEAD instances \cite{wan-optimizers} have throughput of 10 and 50 Mbps whereas virtual firewalls such as Baracuda firewalls \cite{firewalls} have throughput of 100, 200, 400, and 750 Mbps. Hence, to attain a desired throughput in an chain of WAN optimizer and firewall, one has to enumerate all possible combinations of VNF instances for each of the VNFs and choose {the combination} minimizing the demand on physical resources (e.g., CPU cores, memory, etc). Furthermore, the chosen combination of VNF instances has to be placed into the physical machines/hosts in such a way that optimizes the overall bandwidth consumption of the chain. for example, placing a VNF instance far apart from other VNF instances of the same chain will result in increased bandwidth allocation along the path.

These problems are interdependent, and an optimal chain deployment has to solve them all together resulting in a joint optimization problem. Furthermore, a deployment solution should adhere the system implementation aspects regarding distributed deployment of VNF instances, traffic splitting, and accurate load distribution among these instances. Existing optimization models including \cite{BariCNSM15} and \cite{moens2014vnf} assume that the functionality of a middlebox is provided by a single VNF and have not studied this joint optimization problem. In this paper, we address this joint optimization problem by taking into account the system implementation aspects. Specifically, our contribution in this paper are as follows:
\begin{itemize}
    \item {We develop an optimization model to deploy a chain in a distributed and resource efficient manner. Our proposed model abstracts heterogeneity of VNF instances and allows us to deploy a chain with custom throughput without worrying about individual VNF's throughput.} 

    \item {We implement this model using Mixed Integer Programming (MIP) in CPLEX for finding optimal solutions in small scale networks.}

    \item {For larger scale networks, we propose \textit{Kariz}, a local search heuristic, that employs a tuning parameter to balance the speed-accuracy trade-off.}

    \item We evaluate Kariz compared to MIP implementation for various chain-lengths and throughput-demands. The results suggest that Kariz achieves the competitive acceptance ratio of $\sim 80\text{-}100\%$ at an extra cost of less than $25\%$ in comparison to MIP model.
\end{itemize}

The rest of the paper is organized as follows. In \sect{relatedWorks}, we study the related work. \sect{challenges} discusses the system implementation and deployment challenges. We present our problem formulation in \sect{problem}. Our solution is proposed and evaluated in \sect{solution} and \sect{evaluation}, respectively. Lastly, \sect{conclusion} concludes this paper.

%% file: sections/relatedWorks.tex
\section{Related Work}\label{section:relatedWorks}
SFC deals with deployment of VNFs that are chained together to provide a collection of services. CoMb \cite{sekar2012design}, for example, proposes a simplified VNF placement by putting all the VNFs dealing with the same flow on the same fixed physical node (called CoMb box). In contrast, our solution does not restrict VNFs to run on a fixed set of physical nodes, and can be deployed anywhere in the infrastructure.

Bari et al. model a batch deployment of chains, called VNF Orchestration Problem (VNF-OP)\cite{BariCNSM15}. VNF-OP deploys each middlebox in one physical node. VNF-P \cite{moens2014vnf} studies a hybrid scenario of hardware-middlebox and VNFs to provide requested service. None of these models assume that a middlebox is deployed in a  distributed manner. Clayman et al. \cite{clayman2014dynamic} consider the placement of VNFs with respect to several goals, including reducing energy consumption and load balancing. Based on these goals, the best performing algorithm out of \textit{least used host}, \textit{N at a time in a host}, and \textit{least busy host} is chosen.

Sahhaf et al. propose to decompose a chain into more elementary and implementation-close components \cite{serviceChainWithDecomposition}. A selection mechanism determines a decomposition to minimize the mapping cost, and an algorithm deploys the selected decomposition. While this work focuses on the functional decomposition, our goal is to decompose the chain based on performance requirement. In addition, their algorithm does not consider the joint optimization properties of the problem.

The distributed deployment of a chain raises several challenging implementation questions including control plane design, VNF state management, and system abstraction. These challenges have addressed by \cite{splitMerge,gember2012stratos,opennf}. Split/Merge \cite{splitMerge} proposes a system to address challenges of VNF state management and traffic route management. Stratos \cite{gember2012stratos} uses a rather simple technique for both the initial and subsequent placement of VNFs. It packs VNFs that belong to the same chain as close as possible. OpenNF \cite{opennf} supports the idea of packet processing to be redistributed across a collection of VNF instances. However, its focus is to provide a coordinated control plane framework for both internal VNF state and network forwarding state. As such, none of these works consider the optimization problem of deploying a VNF chain in a distributed fashion.

The related works discussed above are compared in \tabl{relatedWorks} in view of four important aspects of the distributed VNF orchestration problem. From the comparison, it is apparent that none of the existing works have considered all the aspects of the optimization problem we study in this paper.

\input{tables/comparison}\label{comp_table} 

%% file: tables/comparison.tex
\begin{table}
    \scriptsize
    \caption{Comparison of Our Work to the Most Related Works}
    \vspace{-0.2cm}
    \setlength\tabcolsep{1pt}
    \begin{tabular}{l|p{1.3cm}|p{1.8cm}|p{1.0cm}|p{1.4cm}|p{1.17cm}}
        Paper   & Methodology  & Distributed VNF deployment & Traffic splitting & Accurate load distribution & Optimal deployment\\
        \hline
        VNF-OP\cite{BariCNSM15}  & Optimization  & \xmark & \xmark & \xmark & \checkmark \\
        Service Dec. \cite{serviceChainWithDecomposition} & Optimization & \xmark & \xmark & \xmark & \checkmark \\
        Split/Merge\cite{splitMerge} & System Imp.        & \checkmark  & \checkmark    & \xmark  & \xmark       \\
        OpenNF \cite{opennf} & System Imp. & \checkmark  & \checkmark & \xmark & \xmark \\
        Our work & Optimization & \checkmark  & \checkmark    & \checkmark  & \checkmark
    \end{tabular}
    \label{table:relatedWorks}
\vspace{-0.6cm}
\end{table}

%% file: sections/challenges.tex
\section{Challenges}\label{section:challenges}
A service chain specifies that the traffic originating from a \textit{source}, is processed by an ordered sequence of middleboxes, and finally is delivered to a \textit{target}. To have transparent underlying VNF instances as well as abstracting the resource requirements of these instances, several system implementation and optimization challenges have to be addressed.


\subsection{System Implementation Challenges}
Middleboxes often operate on data-packets in a \textit{flow} granularity and maintain \textit{state information} on the flows and sessions they process \cite{HILTI, Verdu2008}. The state information consists of configuration and statistical data, and differs from one middlebox to another. By replacing a middlebox with multiple \textit{VNF instances}, the functionality should not change, and these instances have to act unified. Moreover, the traffic processed by a single middlebox, now should be processed by multiple VNF instances. Thus, \textit{consistent state distribution} and \textit{consistent traffic distribution} among the VNF instances are essential.

\subsubsection{Consistent State Distribution}
Deployment of multiple VNF instances to provide functionality of a middlebox requires distribution of the state information. Hence, we need to \textit{model} the state information of middleboxes and \textit{distribute} the state information among the VNF instances consistently. The state information can be classified as \textit{internal} and \textit{external}. The internal state is only stored and used by a single instance, while the external state is distributed and shared across multiple instances. Since the state information is stored in a key-value structure \cite{parrallelNIDS, HILTI, modelingMiddleboxes}, data structures like distributed hash-tables and technologies like Remote Direct Memory Access (RDMA) can fulfill this challenge efficiently. Moreover, it might require to modify the middleboxes to cope with the defined model. There are abstraction models and system implementations that address this challenge. Rajagopalan et al. \cite{splitMerge} introduce a system-level abstraction called Split/Merge that store the internal state exclusively inside each VNF instance, while the external state is distributed and accessible among other instances. As a proof of concept, they implemented FreeFlow as a Split/Merge system, and ported Bro IDS \cite{broIDS} inside it. Further, they analyzed and confirmed the compatability of two other middleboxes, i.e. application delivery controller and stateful NAT64. In addition, Joseph and Stoica \cite{modelingMiddleboxes} provides a model to describe different middleboxes. As concrete examples, firewall, NAT and layer4 and layer 7 load balancer are described using the proposed model. Moreover, Qazi et al. \cite{towardSDNMiddlebox} and OpenNF \cite{opennf} introduce a unified framework to manage the state information.

\subsubsection{Consistent Traffic Distribution}
By replacing a single middlebox with multiple VNF instances, \textit{splitting} and \textit{balancing} the traffic load among these instances are necessary. Per-flow traffic splitting distributes the traffic in granularity of flows, and packets of a flow have to be routed along the same path.
Split/Merge \cite{splitMerge} utilizes a similar approach. However, this approach does not support accurate load distribution and is not always applicable. For instance, if the load of a flow is higher than the throughput of assigned VNF instance, it cannot handle the load and we have to split the traffic to a smaller granularity. \textit{Flowlet switching} \cite{flare, flareOrg, conga} can be leveraged to split the traffic in a more fine-grained granularity. A flowlet is a ``burst of packets from the same flow followed by an idle interval'' \cite{flare}. If the interval between two flowlets is greater than the maximum difference of parallel paths, the second flowlet --and consequaently following flowlets-- can be sent through different paths. Thus, a single flow can split into multiple paths without packet-reordering. Furthermore, accurate load balancing is achieved using short flowlet intervals ($[50, 100]ms$)\cite{flare}. Specifically, flowlets are abundant in data center networks since the latency is very low and the traffic is intensively bursty \cite{bulletTrains2013}. In addition to these distributed methods, the central schemes leveraging SDN and OpenFlow capabilities \cite{laberio} can also be used. For instance, \textit{group tables} \cite{openflowv131} can be used to split and balance the traffic. Combining these schemes with virtualization technologies, such as VXLAN \cite{vxlan} and NVGRE \cite{nvgre} can provide consistent traffic distribution for deployed chains.


We showed the feasibility of distributed deployment of VNF instances to provide the functionality of a middlebox and distributing traffic among these instances. Here, we clearly mention our assumptions to build the ground for our optimization model.
\begin{itemize}
    \item The state information of middleboxes can be classified and distributed among multiple VNF instances.
    \item VNF instances of the same middlebox act as a single unit by accessing the distributed state information.
    \item The host resource overhead of accessing distributed state information is considered in resource demands of VNFs.
    \item Multi-path routing of a single flow among the VNF instances does not alter the functionality of instances as a whole, and shared distributed state information is sufficient for the correct functionality.
    \item The communication overhead to access the distributed state information is negligible compared to the actual service traffic volume.
    \item VNF instances belonging to the same middlebox process the same amount of traffic in similar amount of time.
\end{itemize}

\subsection{Optimization Challenges}\label{section:optimization-challenges}
The optimization challenge is computing an optimal allocation of \textit{host} and \textit{bandwidth} resources to a chain. For each middlebox in a chain, a number of \textit{instances} of each \textit{VNF} are placed to provide the requested throughput. These instances are placed in a set of selected hosts. In addition, the traffic is split and routed among the placed instances. Therefore, following decisions have to be made optimally: \textit{Number of instances} of each VNF, \textit{placement} of these instances in a set of hosts, and \textit{routing the traffic} among the placed instances. These decisions are dependent and need to be made together.

\fig{sample} depicts a deployment of a chain. The substrate network of \fig{network} consists of 6 hosts. Each host has 8 core CPU and 64 GB residual memory. For the sake of simplicity in this example, the switches are not shown, and we assume that presented substrate paths are disjoint. All substrate paths have 130 mbps available bandwidth. The chain of \fig{serviceChainRequest} consists of two VNFs with 210 mbps throughput: an Intrusion Detection System (IDS) and a firewall (FW). The traffic flow comes from host $A$, the source, and after being processed by IDS and FW is sent to host $F$, the target. As listed in \tabl{sample}, there are 4 VNF types for IDS and FW. \fig{deployed} depict the deployed service chain in the network, and \fig{logicalRep} shows the logical representation of this deployment. As shown, three instances for IDS (one $IDS_1$ and two $IDS_2$) and two instances for FW (one $FW_1$ and one $FW_2$) are placed. The IDS instances are installed in hosts $B$ and $D$. The traffic flow splits, and 80 mbps and 130mpbs is routed from the source to hosts $B$ and $D$, respectively. FW instances are installed in hosts $B$ and $E$. In host $B$, the traffic flow after being processed by $IDS_2$ is sent to $FW_1$. Furthermore, $IDS_1$ and $IDS_2$ forward the traffic flow to host $C$ in which instance $FW_2$ is placed. Finally, the traffic flow from the FW instances is sent to the target. Note that it is possible to place the VNF instances in the source and target if there are sufficient available host resources.

\input{figures/sample.tex}

\input{tables/sample.tex}

%% file: figures/sample.tex
\begin{figure*}
    \centering
    \begin{subfigure}[b]{0.09\textwidth}
        \centering
        \includegraphics[width=\textwidth]{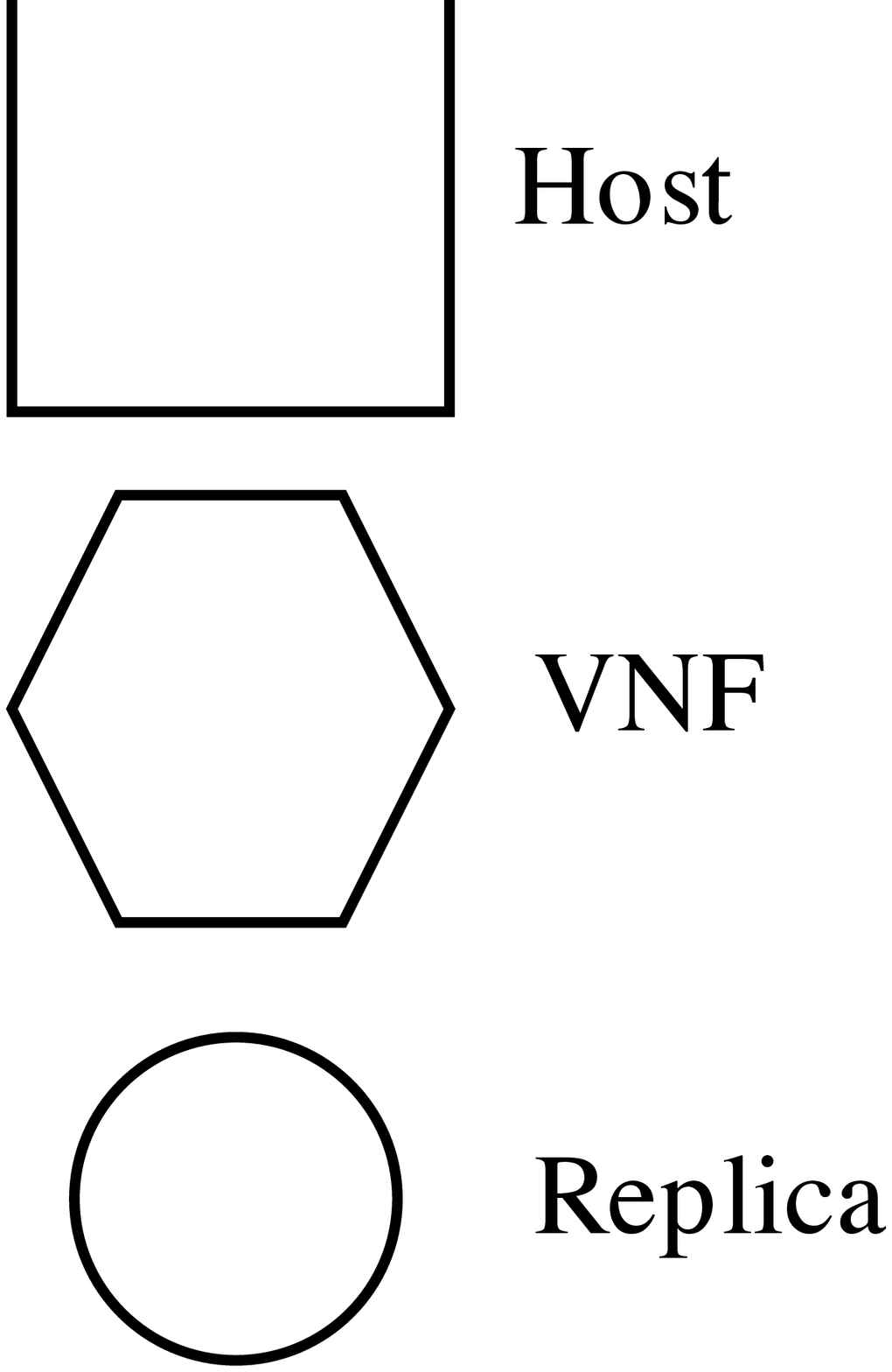}
    \end{subfigure}
    \centering
    \begin{subfigure}[b]{0.21\textwidth}
        \centering
        \includegraphics[width=\textwidth]{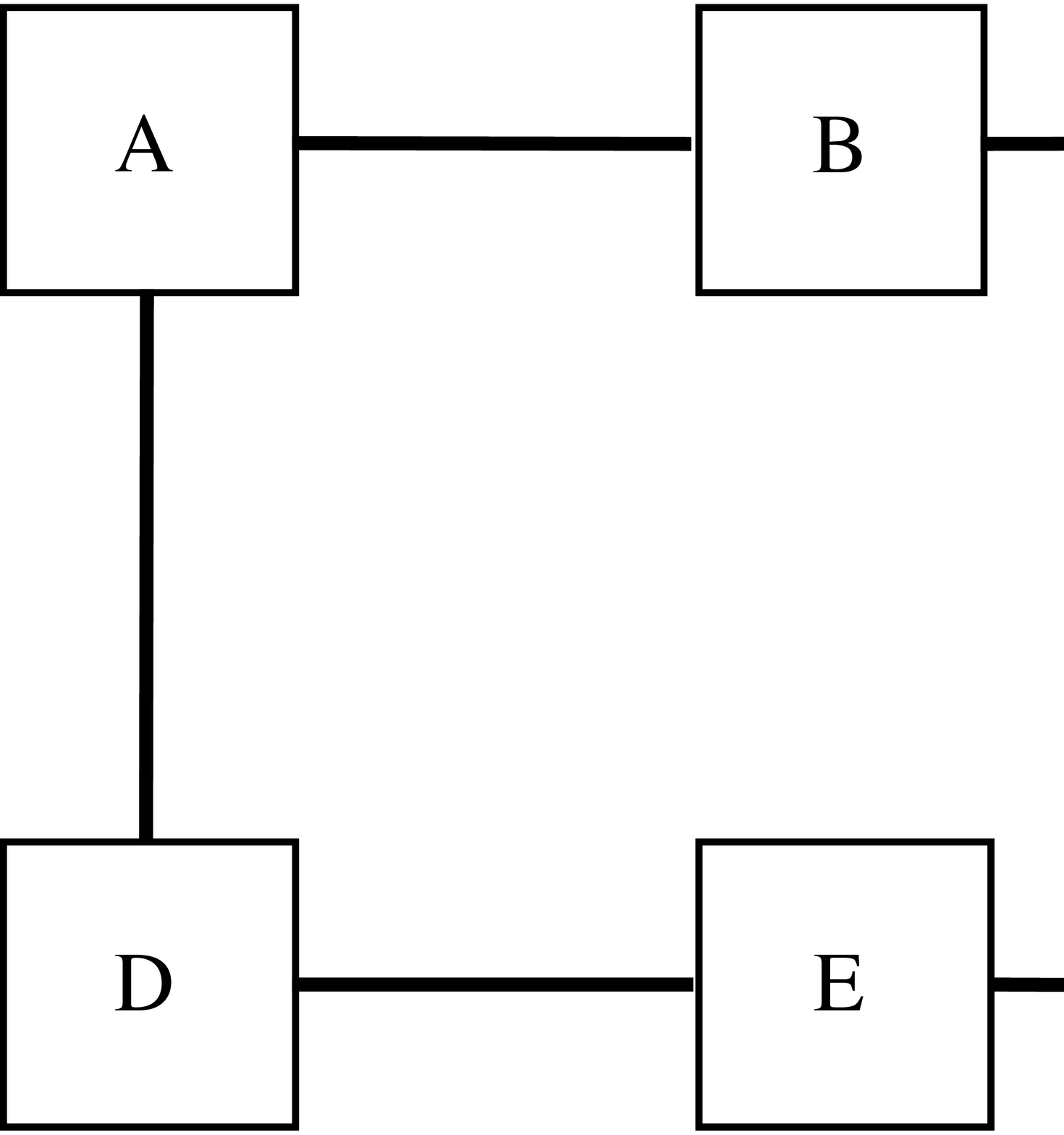}
        \caption{Substrate Network}
        \label{fig:network}
    \end{subfigure}
    \begin{subfigure}[b]{0.21\textwidth}
        \centering
        \includegraphics[width=\textwidth]{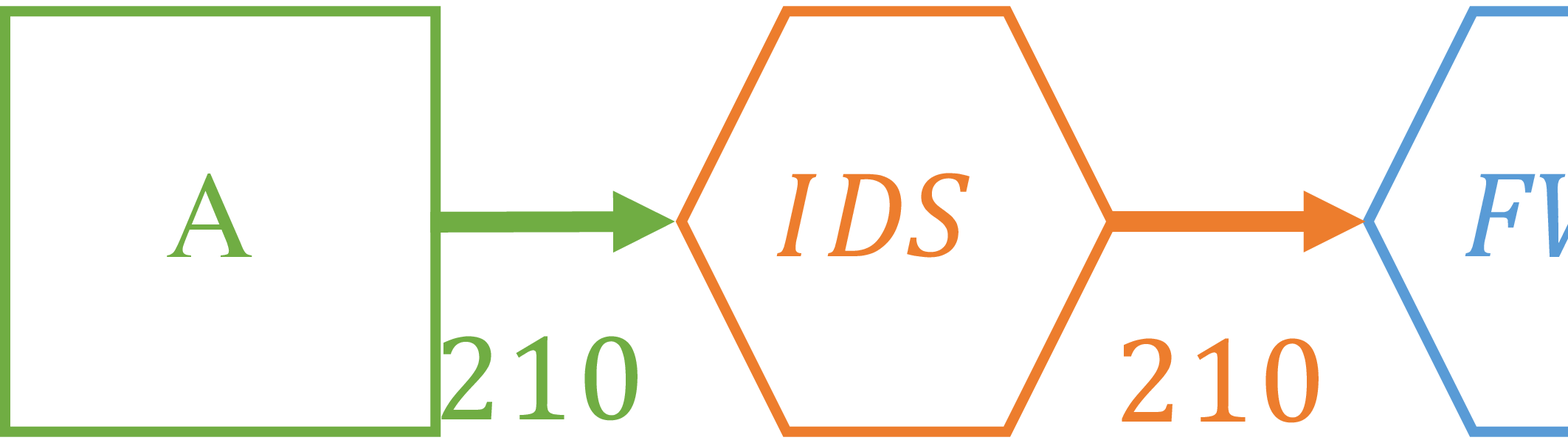}
        \caption{Service Chain}
        \label{fig:serviceChainRequest}
    \end{subfigure}
    \begin{subfigure}[b]{0.21\textwidth}
        \centering
        \includegraphics[width=\textwidth]{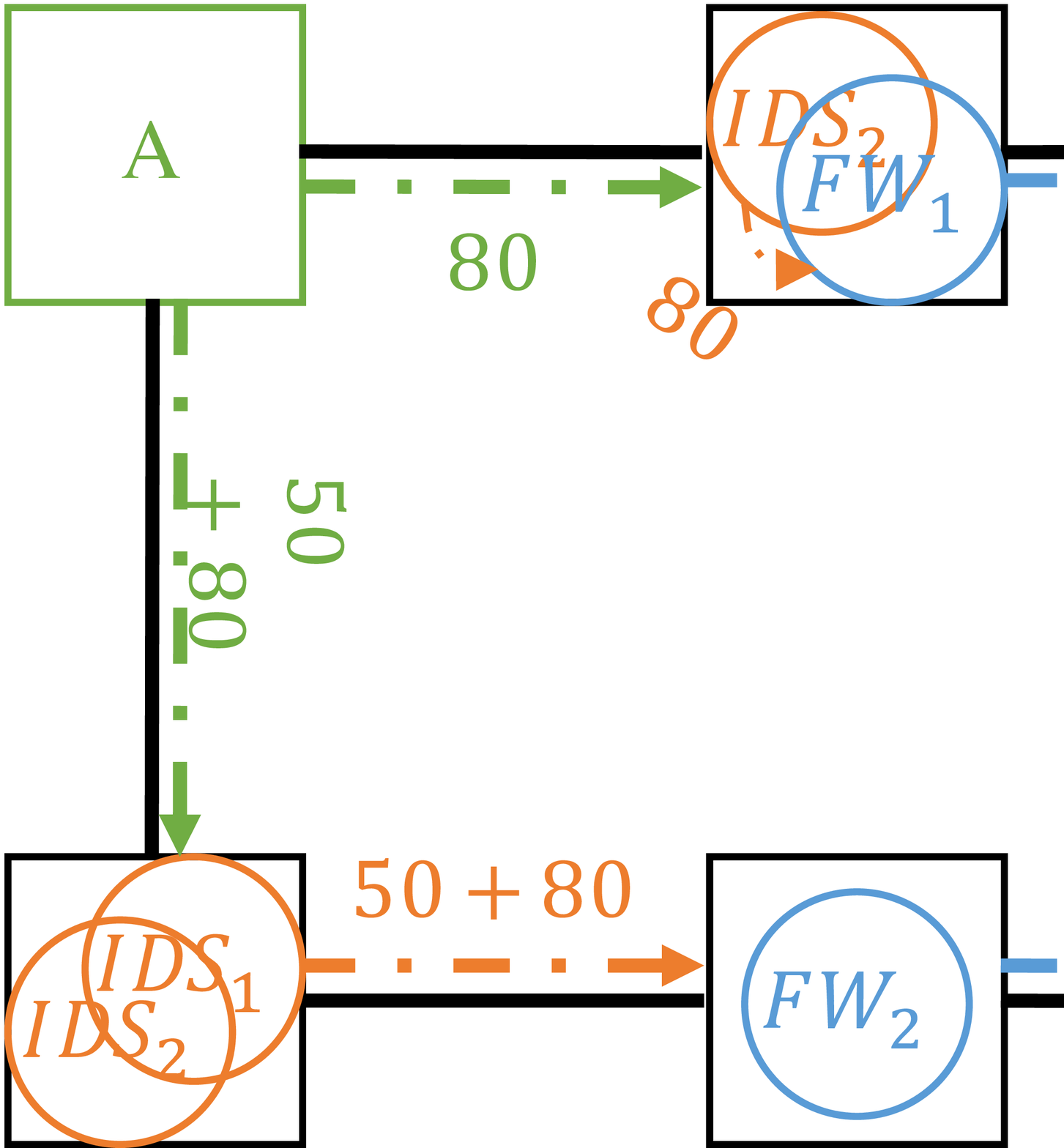}
        \caption{Deployed in Network}
        \label{fig:deployed}
    \end{subfigure}
    \begin{subfigure}[b]{0.21\textwidth}
        \centering
        \includegraphics[width=\textwidth]{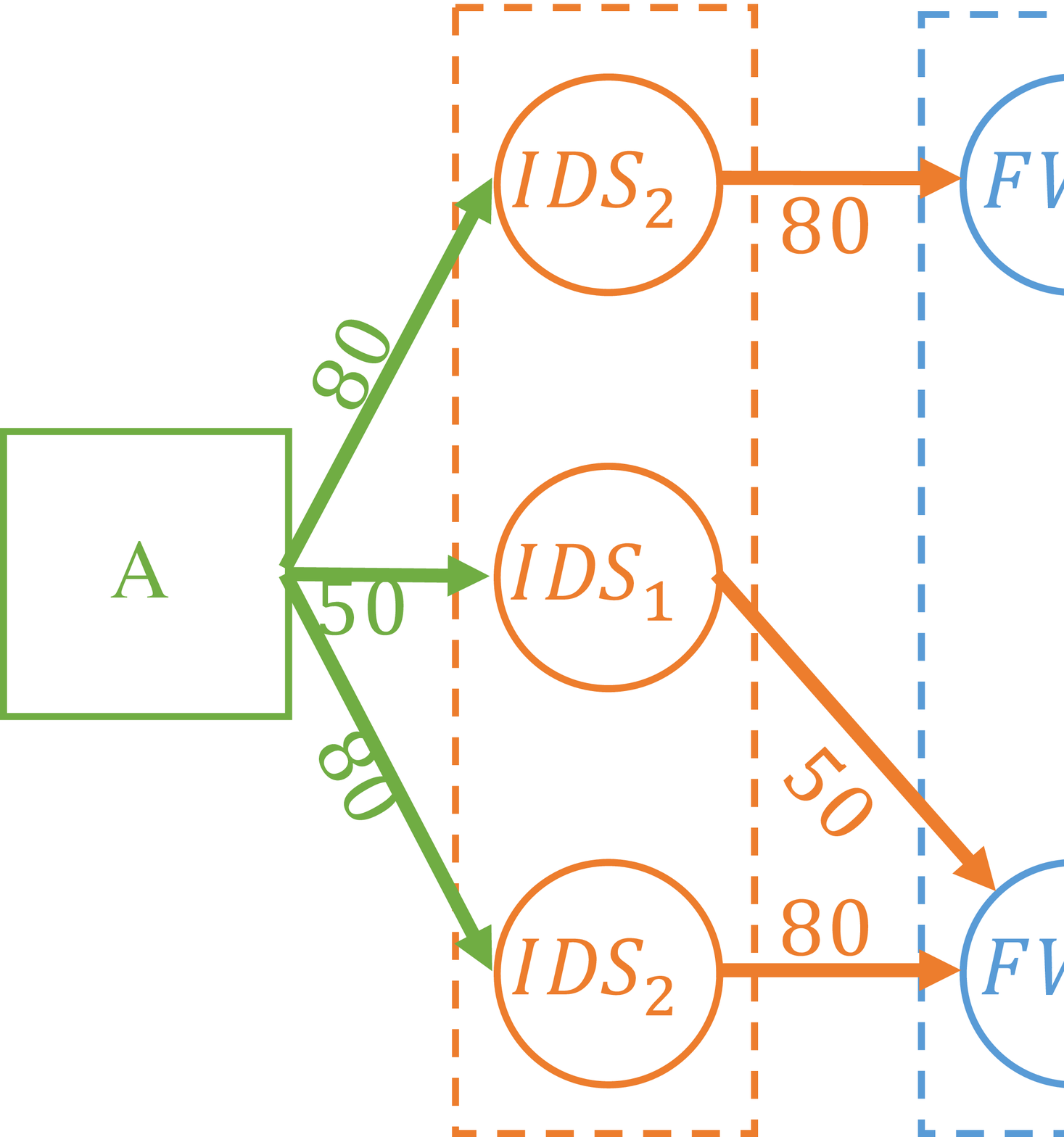}
        \caption{Logical Representation}
        \label{fig:logicalRep}
    \end{subfigure}
    \caption{Deployment of a Service Chain}
    \label{fig:sample}
\vspace{-0.6cm}
\end{figure*}

%% file: tables/sample.tex
\begin{table}
    \caption{VNFs}\vspace{-0.2cm}
    \scriptsize
    \begin{tabular}{c|c|c|c|c}
        Middlebox    & VNF       & Throughput    & CPU demand    & Memory demand   \\\hline
        IDS                 & $IDS_1$   & 50 Mbps       & 1 core        & 24 GB           \\
                            & $IDS_2$   & 80 Mbps       & 1 core        & 32 GB           \\\hdashline
        Firewall            & $FW_1$    & 100 Mbps      & 1 core        & 1.75 GB         \\
                            & $FW_2$    & 200 Mbps      & 2 core        & 3.50 GB         
    \end{tabular}
    \label{table:sample}
\vspace{-0.7cm}
\end{table}

%% file: sections/problem.tex
\section{Service Function Chaining Simplified}\label{section:problem}
Having the assumptions established and optimization challenges discussed, we introduce the formal definitions followed by the mathematical model.

\subsection{Definitions}
\subsubsection{Physical Resources}
$R=\{\text{CPU, memory, storage, \ldots}\}$ represents a set of available physical resources.

\subsubsection{Substrate Network}
Graph $G=(N,E)$ is the substrate network, where $N$ and $E$ are substrate nodes and links, respectively. We use index notation for substrate nodes. For instance, $m < n$ for nodes $m, n \in N$ means that index of $m$ is less than index of $n$. Let $c_{mr} \in \mathbb{R}^+$ denotes the residual capacity of node $m \in N$ for resource $r \in R$. Set $E_m \subseteq E$ represents incident links on node $m$. Moreover, $(m,n) \in E$ is the link between node $m \in N$ and node $n \in N$ and has residual bandwidth capacity of $c_{mn} \in \mathbb{R}^+$.

\subsubsection{Service Chain}
Forwarding graph $\overline{G}=(\overline{N}, \overline{A})$ denotes a chain. We use Service Function (SF) and middlebox synonymously. $\overline{N}$ includes SFs $\overline{V} \subset \overline{N}$, and two endpoints $\overline{s}$ and $\overline{t}$. Traffic flow coming from $\overline{s} \in \overline{N}$, is processed by SFs in the chain, and is forwarded to $\overline{t} \in \overline{N}$. $\overline{s}$ and $\overline{t}$ are the \textit{source} and \textit{target} of the traffic, respectively. Corresponding substrate nodes for source and target are respectively $s \in N$ and $t \in N$. SF $\overline{v}=f(\overline{u})$ is following SF next to SF $\overline{u}$. We define \textit{ring} $(\overline{u},\overline{v}) \in \overline{A}$ as two consecutive SFs $\overline{u}, \overline{v} \in \overline{N}$, where $\overline{v} = f(\overline{u})$. We assume that $\overline{u}$ generates traffic of type $\overline{u}$ and $\overline{v}$ consumes this traffic type. Each ring ${(\overline{u},\overline{v})} \in \overline{A}$ has the \textit{throughput demand} of $\overline{b}$ representing the integer volume of traffic flow that is generated or consumed by the ring nodes.

\subsubsection{VNFs}
Set $V$ denotes VNFs. Each VNF $u \in V$ has throughput $q_u \in \mathbb{R}^+$ showing the maximum traffic volume that $u$ can process. Besides, $d_{ur} \in \mathbb{R}^+$ is the demand of $u$ for resource $r \in R$. For $\overline{s}, \overline{t} \in \overline{N}$, we assume there are VNFs $u_{\overline{s}} \in \overline{V}$ and $u_{\overline{t}} \in \overline{V}$, respectively. These VNFs have throughput of $\overline{b}$ and no demand for any resource. Finally, VNFs of type $\overline{u} \in \overline{V}$ are identified by $V_{\overline{u}}$.

\subsection{Mathematical Model}
\subsubsection{Decision Variables}
$x^{\overline{u}}_{mn} \in \mathbb{R}$ is the volume of traffic of type $\overline{u} \in \overline{N} / \{\overline{t}\}$ on substrate link $(m,n) \in E$. Target $\overline{t}$ is excluded from this definition because it only consumes the traffic, therefore no traffic of this type exists in the network. Variable $y_{mu} \in \mathbb{Z}$ is the number of instances of VNF $u \in V$ in substrate node $m \in N$. VNF instances of $V_{\overline{u}}$ installed in node $m \in N$ provide throughput of type $\overline{u} \in \overline{N}/\{t\}$. Decision variable $z_{m\overline{u}} \in \mathbb{R}$ denotes the allocated throughput of these VNF instances. A solution for the problem is represented by a tuple of allocation vectors $(X, Y, Z)$ which are defined as follows. Let vector $X_{\overline{u}}=\{x_{mn}^{\overline{u}} : \forall(m,n) \in E\}$ be allocated bandwidth of links to traffic of type $\overline{u}$, and $X=\bigcup_{\overline{u}\in \overline{N}/\{t\}}X_{\overline{u}}$. If $Y_{\overline{u}}=\{y_{mu} : \forall m \in N, \forall u \in V_{\overline{u}}\}$ identifies the VNF instantiated for SF $\overline{u} \in \overline{N}$, let $Y=\bigcup_{\overline{u}\in \overline{N}/\{t\}}Y_{\overline{u}}$. Finally, $Z_{\overline{u}}=\{z_{m\overline{u}} : \forall m \in N\}$ denotes allocated throughput of type $\overline{u} \in \overline{N}/\{t\}$ in every node, and $Z=\bigcup_{\overline{u}\in \overline{N}/\{t\}}Z_{\overline{u}}$.

\subsubsection{Substrate Node Capacity Constraint}
\eqn{substrateNodeCapConst} guarantees the resource capacities of substrate nodes in which instances are placed are respected.
\input{equations/substrateNodeCapConst}

\subsubsection{Location Constraint}
Equalities in \eqn{locationConst} ensure that a instance of $u_{\overline{s}}$ and a instance of $u_{\overline{t}}$ are only placed in $s \in N$ and $t \in N$, respectively.
\input{equations/locationConst}

\subsubsection{Substrate Link Capacity Constraint}
\eqn{substrateLinkCapConst} makes sure that the capacities of substrate links are not violated.
\input{equations/substrateLinkCapConst}

\subsubsection{Throughput Constraint}
\eqn{reThroughputConst} assures that the aggregate throughput capacity of instances of VNFs of type $\overline{u} \in \overline{N}$ placed in substrate node $m \in N$ is more than allocated throughput $z_{m\overline{u}}$.
\input{equations/reThroughputConst}

\subsubsection{Throughput Demand Constraint}
\eqn{reBandwidthDemandCons} guarantees that for each SF $\overline{u} \in \overline{V}$, throughput of $\overline{b}$ is allocated by VNF instances of $V_{\overline{u}}$.
\input{equations/reBandwidthDemandCons}

\subsubsection{Flow Conservation Constraint}\label{section:flowConservationConstraint}
\eqn{flowConsConst} is the modified version of flow conservation constraint \cite{tomlin1966mcmf}. Let say in node $m \in N$, VNF instances of types $\overline{u}$ and $\overline{v} = f(\overline{u})$ are installed. Therefore, VNF instances of $V_{\overline{v}}$ locally process a volume of traffic of type $\overline{u}$ generated by instances of $V_{\overline{u}}$. This volume is $z_{m\overline{v}}$. Not processed traffic volume should comes outside the node $m$. This constraint assures this phenomenon.
\input{equations/flowConsConst}

\subsubsection{Bandwidth Allocation Cost}
\eqn{bandwidthCost} denotes the bandwidth resource allocation cost. Coefficient $\beta \in \mathbb{R}^+$ identifies the relative importance of bandwidth resources. Analogously, $B(X_{\overline{u}})$ is the bandwidth cost for SF $\overline{u}$.
\input{equations/bandwidthCost}

\subsubsection{Host Resource Allocation Cost}
\eqn{hostCost} is the cost of allocating host resources to place VNF instances. $\alpha_r \in \mathbb{R}^+$ is a coefficient denoting the relative importance of resource $r \in R$. Similarly, $H(Y_{\overline{u}})$ and $H(y_{mu})$ represent this cost for SF $\overline{u} \in \overline{V}$ and VNF $u \in V$, respectively.
Note that we can compute cost of $H(y_{mu})$ if $z_{m\overline{u}}$ is given\footnote{By solving a variant of knapsack problem as explained in \sect{routeAndVNFInstances}}. Let $H(z_{m\overline{u}})$ be this computed cost.
\input{equations/hostCost}

\subsubsection{Objective Function}
\eqn{objectiveFunction} is minimization of aggregate cost of allocating host and bandwidth resources.
\input{equations/objectiveFunction}

This problem is NP-Hard. Even if the number of instances and throughput allocations for every VNF are known, the problem still generalizes the NP-Hard problem of \textit{virtual network embedding problem with path splitting} \cite{Yu2008RVN, chowdhury2009}. Due to intractability of the problem for larger scales, we introduce a heuristic which approximates the optimal solution in a reasonable time. 

%% file: equations/substrateNodeCapConst.tex
\begin{equation}
    \forall m \in N:
        \forall r \in R:
            \sum_{u \in V} {
                y_{mu} d_{ur}
                \leq
                c_{mr}
            }
    \label{equation:substrateNodeCapConst}
\end{equation}

%% file: equations/locationConst.tex
\begin{equation}
    \begin{aligned}
        y_{s u_{\overline{s}}} &= 1 
        ,&
        \sum_{m \in N / \{s\}} {
            y_{m u_{\overline{s}}}
        } = 0
        \\
        y_{t u_{\overline{t}}} &= 1
        ,&
        \sum_{m \in N / \{t\}} {
            y_{m u_{\overline{t}}}
        } = 0
    \end{aligned}
    \label{equation:locationConst}
\end{equation}

%% file: equations/substrateLinkCapConst.tex
\begin{equation}
    \forall (m,n) \in E, m < n:
        \sum_{\overline{u} \in \overline{N}} {(
            x^{\overline{u}}_{mn} +
            x^{\overline{u}}_{nm}
        )}
        \leq
        c_{mn}
    \label{equation:substrateLinkCapConst}
\end{equation}

%% file: equations/reThroughputConst.tex
\begin{equation}
        \forall m \in N:
        \forall \overline{u} \in \overline{N}:
            \sum_{u \in V_{\overline{u}}}{
                y_{mu}q_u
            }
            \geq
            z_{m\overline{u}}
\label{equation:reThroughputConst}
\end{equation}

%% file: equations/reBandwidthDemandCons.tex
\begin{equation}
    \forall \overline{u} \in \overline{N}:
        \sum_{m \in N} {
            z_{m \overline{u}}
        }
        = 
        \overline{b}
    \label{equation:reBandwidthDemandCons}
\end{equation}

%% file: equations/flowConsConst.tex
\begin{equation}
    \begin{aligned}
         \forall m \in N: \forall \overline{u} \in \overline{N} / \{\overline{t}\}: \overline{v} = f(\overline{u}): \\
            \sum_{(m,n) \in E_m} {\big(
                x^{\overline{u}}_{mn} -
                x^{\overline{u}}_{nm}
            \big)}
            =
            \big(
                z_{m \overline{u}} - z_{m \overline{v}}
            \big)
    \end{aligned}
    \label{equation:flowConsConst}
\end{equation}

%% file: equations/bandwidthCost.tex
\begin{equation}
    B(X) = 
        \sum_{\overline{u} \in \overline{N} / \{t\}} {
            \sum_{(m,n) \in E} {
                \beta
                    x_{mn}^{\overline{u}}
            }
        }
    \label{equation:bandwidthCost}
\end{equation}

%% file: equations/hostCost.tex
\begin{equation}
    H(Y) =
        \sum_{u \in V}{
            \sum_{r \in R} {
                \alpha_{r} d_{ur} y_{mu}
            }
        }
    \label{equation:hostCost}
\end{equation}

%% file: equations/objectiveFunction.tex
\begin{equation}
    \min
    \Big(
        B(X) + H(Y)
    \Big)
    \label{equation:objectiveFunction}
\end{equation}

%% file: sections/solution.tex
\section{Kariz: Heuristic Solution}\label{section:solution}
Before explaining our solution, we construct a visualization tool to simplify our description. Let assume that each $\overline{u} \in \overline{N}$ is deployed in a \textit{layer}. Each layer contains a set of nodes in which VNF instances of corresponding type can be installed. In other words, in the layer corresponding to $\overline{u}$, we initially place nodes in which at least a VNF $v \in V_{\overline{u}}$ can be instantiated. More precisely, this layer is a subset of nodes and is denoted by $L(\overline{u})$. \fig{layersInitial} depicts the layers for chain. As shown in \fig{layersInitial}, $s$ and $t$ are the only present nodes in layers $L(\overline{s})$ and $L(\overline{t})$, respectively. Further, nodes $\{s,m\}$ and $\{n,t\}$ are respectively included in layers $L(\overline{u})$ and $L(\overline{v})$ because these nodes have sufficient resource to host VNF instances of these SFs. We can now describe our problem as the problem of routing between layers to bring the traffic from the first layer $L(\overline{s})$ to last layer $L(\overline{t})$. In each layer $L(\overline{u})$, traffic passes through a set of nodes in which VNF instances of $V_{\overline{u}}$ are placed. \fig{layersWithFlows} presents a sample solution for the chain of \fig{layersSCR}.

\input{figures/layers.tex}

Inspired by \cite{guha2000hierarchical, pal2001facility}, we develop a local search heuristic, \textit{Kariz}, which routes traffic layer by layer. We provide the process first, and then explain an overview of the details. Kariz is shown in \alg{solution} and works as follows. At the beginning, we set initial solution as empty (line 1). Starting from layer $L(\overline{s})$ (line 2), iteratively route $\overline{b}$ volume of traffic from layer $S=L(\overline{u})$, \textit{source-layer}, to next layer $T=L(\overline{v})$, \textit{sink-layer} (lines 3-11). After finding the optimal route between two layers (line 5), compute the number of VNF instances of $V_{\overline{v}}$ by considering the allocated throughput (line 6). Add the solution of sink-layer to the earlier solution (line 7). Improve the current solution (line 8), and update layers (line 9). Now, traffic has reached the sink-layer; consider this layer as new source-layer (line 10). Repeat this procedure if traffic has not reached the last layer yet, and there are nodes in new source-layer (line 11).
\input{algorithms/solution.tex}

Yet, we have not clarified how the routing between two layers and the number of VNF instances in the sink-layer are computed; how the solution is improved; also how the layers are updated.

\subsection{Route and VNF Instances}
\label{section:routeAndVNFInstances}
Function $route(.)$ in \alg{solution} computes the route between two layers by solving the multi-source multi-sink Minimum Cost Flow Problem (MCFP)\cite{mcf1989}. MCFP is the problem of routing a volume (say $\overline{b}$) of a \textit{commodity} (in our case traffic of type $\overline{u}$) from multiple sources (say source-layer) to multiple sinks (in our case sink-layer). Any multi-source multi-sink MCFP can be modeled as a single-source single-sink MCFP which is solvable in polynomial time \cite{mcf1989}. For our problem, this is achieved by representing the source- and sink-layers with imaginary nodes \textit{super-source} and \textit{super-sink}, respectively. \fig{mcf} depicts this model for layers $S$ and $T$ in \fig{layers}. The procedure is as follows. Add super-source and connect it to every node $m \in S$ in the source-layer with a directed-link whose capacity is $z_{s\overline{u}} \in Z$. For the sink-layer, add super-sink node and connect every node $n \in T$ using a directed-link. The capacity of the directed-link connecting node $n$ to super-sink is the maximum throughput $\max(z_{n\overline{v}})$ of the VNF instances that can be installed in node $n$. There is no cost to send the traffic via these links. As the result, the minimum cost route of traffic from super-source to super-sink gives the optimal routing between the two layers. If $p$ represents super-sink, the throughput allocation in each $n \in L(\overline{v})$ is $z_{n\overline{v}}=x_{np}^{\overline{u}}$.

Finding the capacity of directed-links from sink-layer to the super-sink is similar to the problem of function $vnf\text{-}instances(.)$. Former is finding the maximum throughput $\max(z_{n \overline{v}})$ out of VNF instances that can be installed in node $n$. Latter is finding the minimum allocation of resources to VNF instances providing throughput of at least $z_{n\overline{v}}$ in each node $n \in L(\overline{v})$. In fact, these two problems are \textit{dual} and can be modeled as a \textit{multidimensional knapsack problem} \cite{lin1998biblographical}.
Think of the node as $|R|$-\textit{dimensional knapsack}, each \textit{dimension} corresponding to a resource $r \in R$. The \textit{items} to be packed are VNF instances with \textit{profits} of their throughputs and \textit{weights} of their host resources demands. Although this problem is known to be NP-Hard \cite{lin1998biblographical}, since the resources of a single physical machine, specially number of CPU cores are limited, and the problem size is small. Hence, we can solve it efficiently. Alternatively, as CPU cores are the most expensive and restricted resources, a feasible solution optimizing the number of allocated cores is a good optimum.

\input{figures/mcf.tex}

\subsection{Solution Improvement Rounds}\label{section:solImprove}
Routing of traffic between two layers might result in fragmented host resource allocation with high cost. Therefore, we need to improve the solution. Function $improve(.)$ as presented in \alg{improve} facilitates this: Repeatedly search for some \textit{actions} to improve the solution (lines 2-8). If no such action is found, report the current solution (line 4-6). Otherwise, perform the action with greatest drop in the cost, the best \textit{admissible} action (line 7), and continue with the adjusted solution. We define actions and \textit{admissibility} in \sect{actions} and \sect{sufficientImprovements}, respectively.

\input{algorithms/improve.tex}

\subsubsection{Actions}
\label{section:actions}
An action is a \textit{local transformation} intended to reduce the solution cost. Let $(X^{'}, Y^{'}, Z^{'})$ be the modified solution after performing an action on a current solution $(X, Y, Z)$. The cost difference before and after performing an action is regarded as the \textit{action cost}, as defined in \eqn{actionCost}. The best action has the lowest cost.

\input{equations/actionCost}

We define the following actions that are variants of actions used by \cite{pal2001facility}:
\begin{itemize}
    \item $add(n,L(\overline{v}),\delta)$:
    Include node $n \in N$ in $L(\overline{v})$ and allocate more $\delta > 0$ units of throughput in this node ($z_{n\overline{v}} \gets z_{n\overline{v}} + \delta$). Then, find the minimum cost routing from layer $L(\overline{v})$ to next and previous layers in the current solution, given allocated throughputs of $L(\overline{v})/\{n\}$. The next and previous layers are $L(\overline{w})$ and $L(\overline{u})$ if $\overline{w}=f(\overline{v})$ and $\overline{v}=f(\overline{u})$, respectively. Finally, tune the allocated throughput of nodes $L(\overline{v})$. This action is shown in \fig{add}.

    \item $open(n,M,L(\overline{v}), \delta)$:
    Add node $n \in N$ into layer $L(\overline{v})$, remove nodes $M \subseteq L(\overline{v})$, and allocate more $\delta > 0$ units of throughput in node $n$ ($z_{n\overline{v}} \gets z_{n\overline{v}} + \delta$). Finally, reroute the traffic either received or originated in layer $L(\overline{v})$. This action replaces a set of fragmented VNFs installed in different nodes $M$ with VNFs collocated in one node $n$. This action makes sense only if $\delta \geq \sum_{m \in M}(z_{m\overline{v}})$. We used a similar action, $install$, in elastic placement of VNFs in \cite{ghaznaviCloudNet15}. \fig{open} depicts this action.
\end{itemize}

\input{figures/actions.tex}

Traffic routing in the above actions is a bit different from routing in function $route(.)$. The difference is routing of two different traffic types. Still this problem is tractable, and we can model it as a multi-commodity MCFP that is solvable in polynomial time.

We also need to examine actions and select the best in polynomial time and ensure that the number of performed actions is not exponential. Particularly, we need to select the best action with sufficient improvement efficiently. These criteria, \textit{efficient action selection} and \textit{sufficient improvement}, are essential to assure that the algorithm terminates in polynomial time.

\subsubsection{Efficient Action Selection}
The number of possible $add(n,L(\overline{v}),\delta)$ actions are at most $|N| \times |\overline{V}| \times \overline{b}$ under the assumption of integrality of $\overline{b}$. Hence, it is possible to check all actions and select the best one in polynomial time. We can even do better and select the value of $\delta$ by considering the throughputs of VNFs $V_{\overline{v}}$. However, number of possible $open(.,M, L(\overline{v}), .)$ actions can be exponential because of the large number of possible subsets $M \subseteq L(\overline{v})$. Thus, we need an efficient procedure to select a good $open(.)$ action. For a fixed layer $L(\overline{v})$, fixed node $n \in N$ and fixed $\delta$, we find this subset in a greedy procedure working as follows. Starting from empty set $M$, iteratively remove a node $m$ from $L(\overline{v})$ and add it to $M$. Removing this node has the minimum cost vs. other nodes $L(\overline{v}) / m$. Continue this procedure while such a node $m \in L(\overline{v})$ exists, the removal of $m$ decreases the cost, and $m$'s throughput is less than $\delta -\sum_{m \in M}z_{m\overline{v}}$. This procedure repeatedly removes an individual node $m \in L(\overline{v})$ whose removal produces the highest decrease in both bandwidth and host resource allocation costs.

\subsubsection{Sufficient Improvement} \label{section:sufficientImprovements}
Still the number of actions can be large due to exponential number of performed actions with minor improvement. To solve this problem, only actions with sufficient improvement of the cost are applied. An action with sufficient improvement is called \textit{admissible}. More precisely, we define an action as admissible if it improves the solution no less than $\frac{\epsilon}{4|N|}\big(B(X) + H(Y)\big)$ for some tuning parameter $\epsilon > 0$ \cite{mahdian2003universal}. Using $\epsilon$, we can control the trade-off between accuracy and speed of our solution. Let $(X^*, Y^*, Z^*)$ be the optimal solution. Since the optimal solution is the lower bound for our solution, the number of performed actions will be at most $\frac{4|N|}{\epsilon}\ln{\frac{B(X) + H(Y)}{B(X^*) + H(Y^*)}}$.

\subsection{Update Layers}
As the last piece of the puzzle, function $update\text{-}layers(.)$ updates the nodes in every layer. From a layer $L(\overline{u})$ to which traffic is already reached, every node $m \in L(\overline{u})$ is eliminated if this node does not allocate throughput of type $\overline{u}$. From other layers, nodes whose resources are allocated and hereafter cannot host corresponding VNF instances are excluded. Layers $L(\overline{s})$ and $L(\overline{t})$ are kept out of the update.

%% file: figures/layers.tex
\begin{figure}
    \centering
    \begin{subfigure}[b]{0.21\textwidth}
        \centering
        \includegraphics[width=\textwidth]{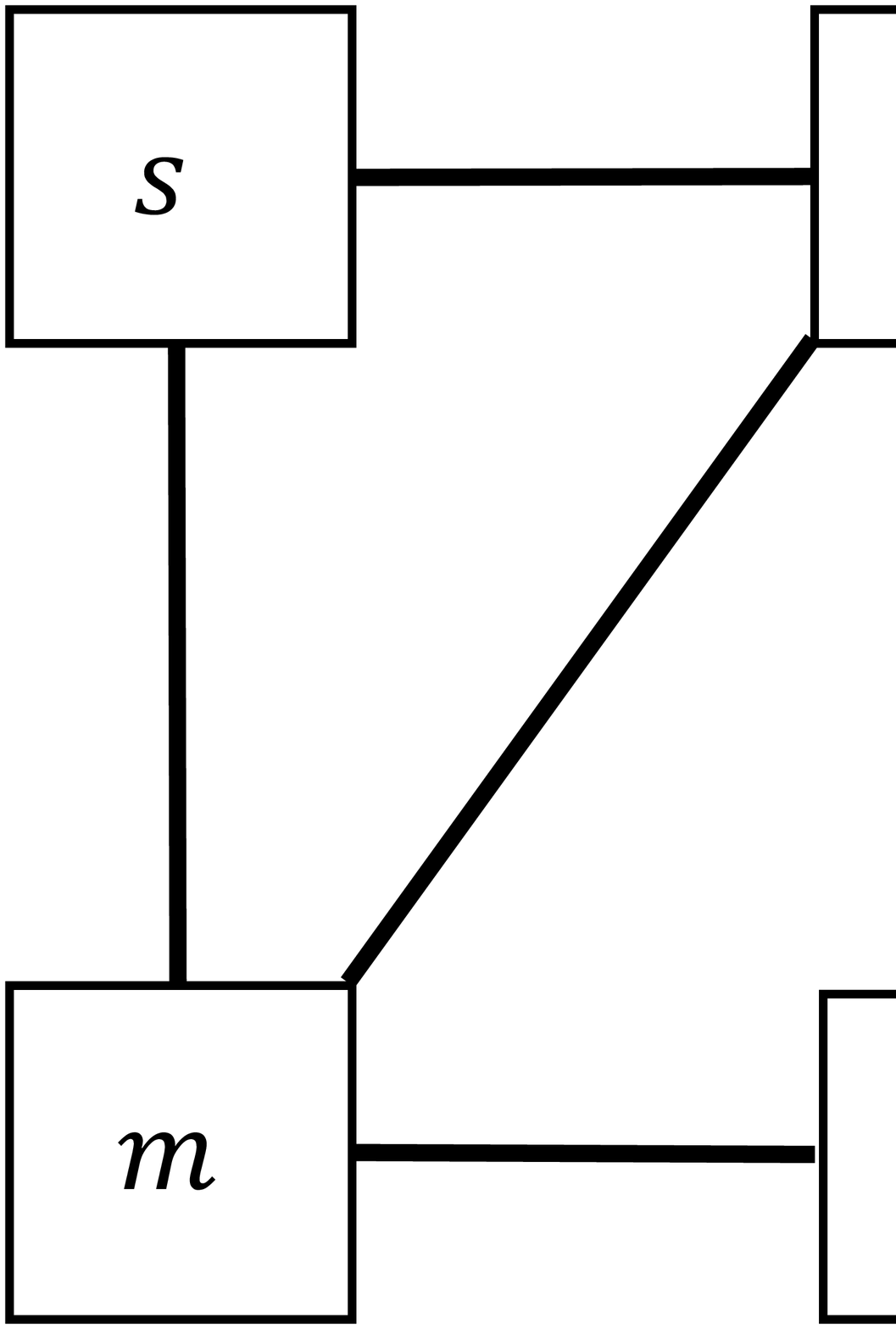}
        \caption{Substrate Network}
        \label{fig:layersNetwork}
    \end{subfigure}
    \begin{subfigure}[b]{0.21\textwidth}
        \centering
        \includegraphics[width=\textwidth]{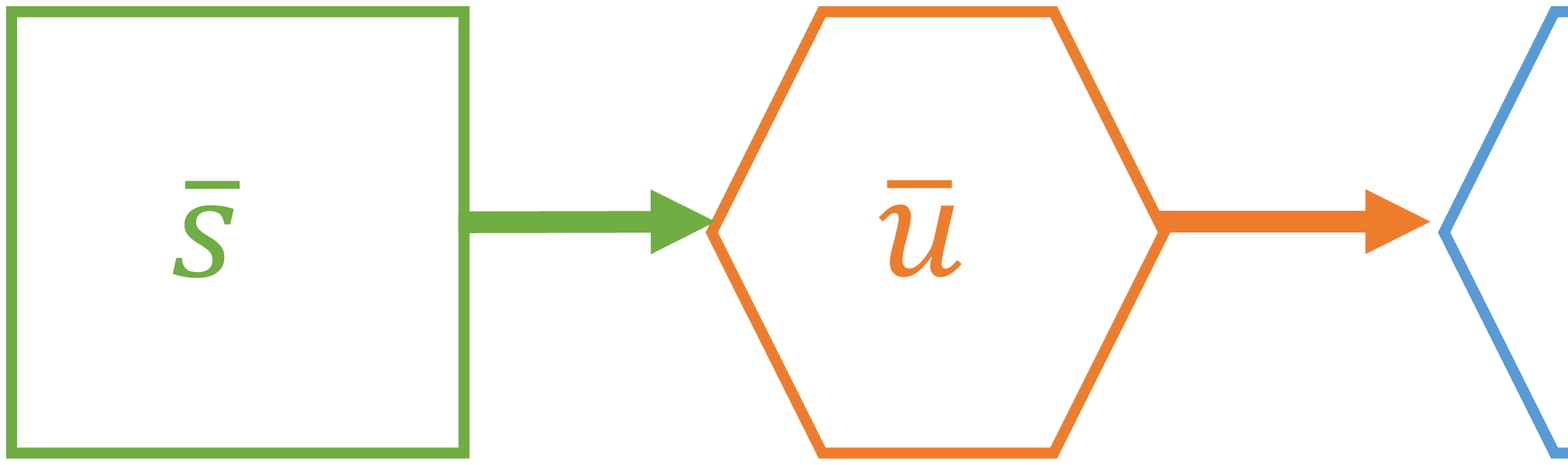}
        \caption{Service Chain}
        \label{fig:layersSCR}
    \end{subfigure}
    \begin{subfigure}[b]{0.21\textwidth}
        \centering
        \includegraphics[width=\textwidth]{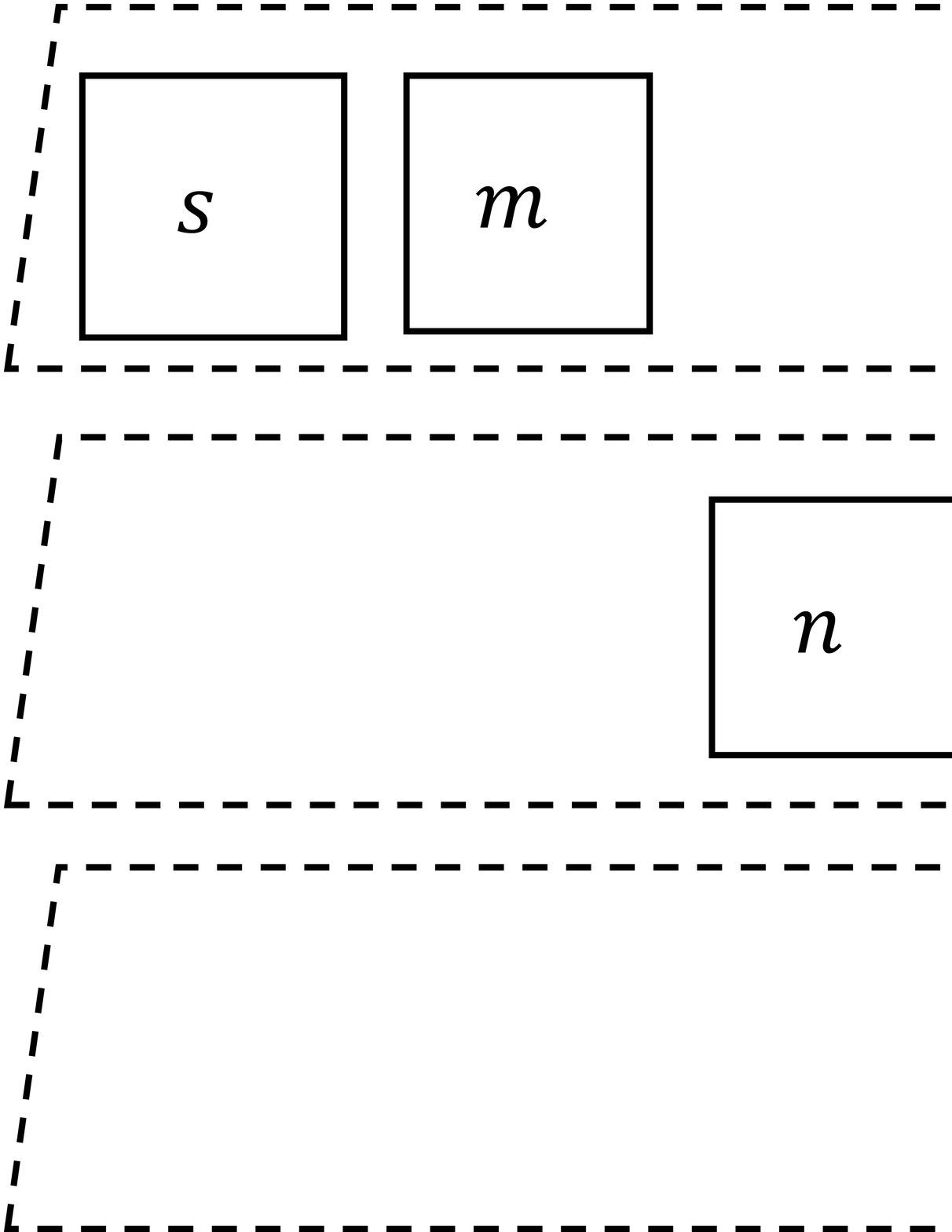}
        \caption{Layers}
        \label{fig:layersInitial}
    \end{subfigure}
    \begin{subfigure}[b]{0.21\textwidth}
        \centering
        \includegraphics[width=\textwidth]{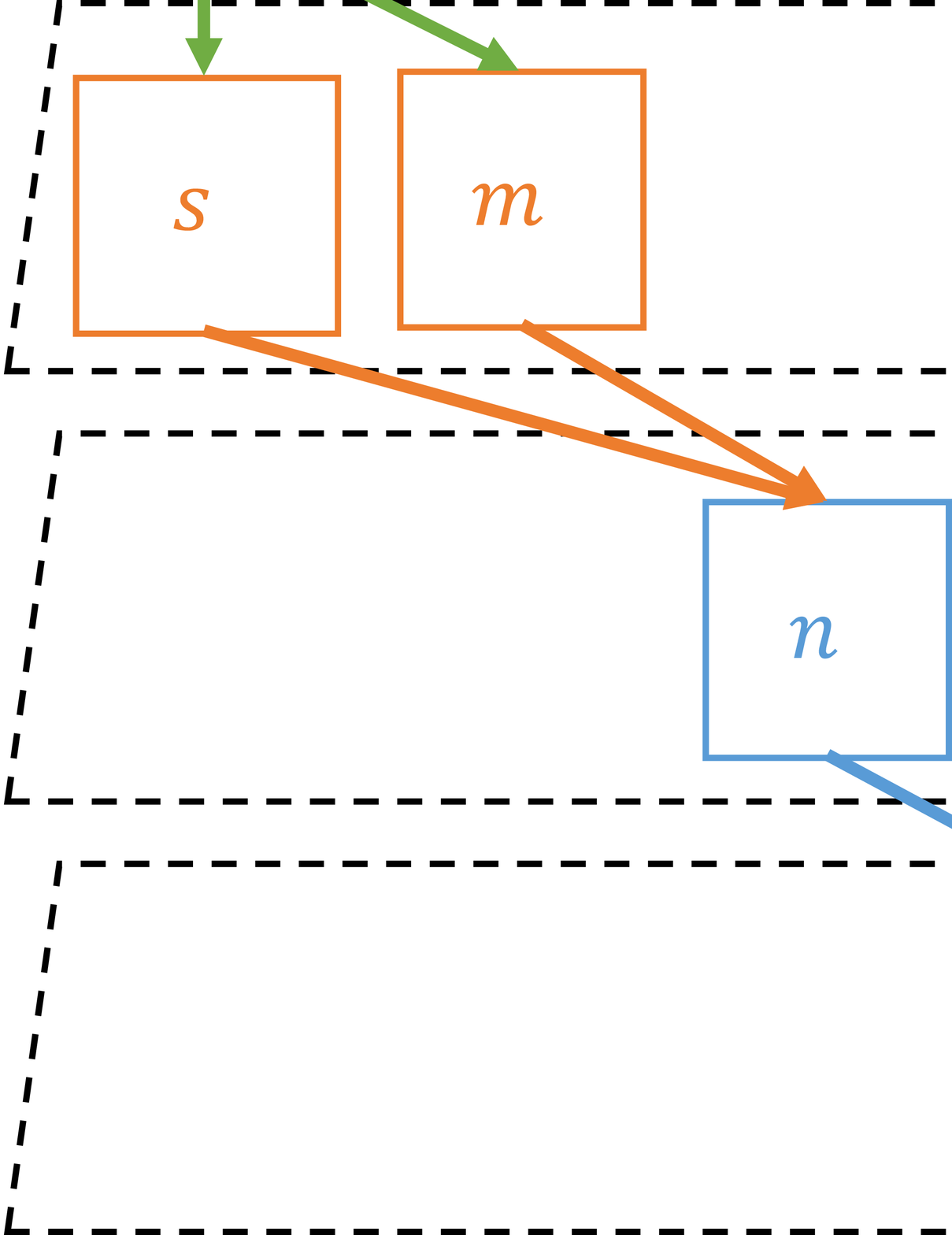}
        \caption{A Sample Solution}
        \label{fig:layersWithFlows}
    \end{subfigure}
    \caption{Layers}
    \label{fig:layers}
\vspace{-0.6cm}
\end{figure}

%% file: algorithms/solution.tex
\begin{algorithm}
\small
\caption{Kariz Algorithm}
\label{algorithm:solution}
	\begin{algorithmic}[1]
        \State{$(X, Y, Z) \gets (\emptyset, \emptyset, \emptyset)$;}
        \State{$\overline{u} \gets \overline{s}$;
               $z_{s\overline{s}} \gets \overline{b}$;
               $z_{t\overline{t}} \gets \overline{b}$;
               $S \gets L(\overline{s})$;}
        \Do
            \State{$\overline{v} \gets f(\overline{u})$;
    	           $T \gets L(\overline{v})$;}
    	    \State{$X_{\overline{v}}, Z_{\overline{v}} \gets route(S, T, \overline{b})$;}
            \State{$Y_{\overline{v}} \gets vnf\text{-}instances(Z_{\overline{v}})$;}
            \State{$(X, Y, Z) \gets (X \cup X_{\overline{v}}, Y \cup Y_{\overline{v}}, Z \cup Z_{\overline{v}})$;}
            \State{$improve(X, Y, Z)$;}
            \State{$update\text{-}layers(Y)$;}
            \State{$\overline{u} \gets \overline{v}$;
                   $S \gets L(\overline{v})$;}
        \doWhile{\big($\overline{u} \neq t$ and $S \neq \emptyset$\big);}
	\end{algorithmic}
\end{algorithm}

%% file: figures/mcf.tex
\begin{figure}
    \centering
    \includegraphics[width=0.24\textwidth]{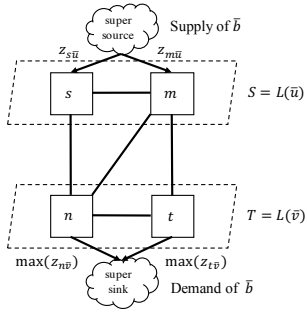}
    \caption{Routing as Single-Source Single-Sink MCFP}
    \label{fig:mcf}
\vspace{-0.6cm}
\end{figure}

%% file: algorithms/improve.tex
\begin{algorithm}
\small
\caption{Function $improve(.)$}
	\label{algorithm:improve}
	\begin{algorithmic}[1]
	\Function{$improve$}{$X$, $Y, Z$}
	    \Loop
	        \State{$a \gets best\text{-}action(X, Y, Z)$;}
	        \If{not $admissible(a)$}
	            \State{return $(X, Y, Z)$;}
	        \EndIf
	        \State{$perform\text{-}action(X, Y, Z, a)$;}
	    \EndLoop
	\EndFunction
	\end{algorithmic}
\end{algorithm}

%% file: equations/actionCost.tex
\begin{equation}
    \big(
        B(X^{'}) + H(Y^{'})
    \big) -
    \big(
        B(X) + H(Y)
    \big)
    \label{equation:actionCost}
\end{equation}

%% file: figures/actions.tex
\begin{figure}
    \centering
    \begin{subfigure}[b]{0.21\textwidth}
        \centering
        \includegraphics[width=\textwidth]{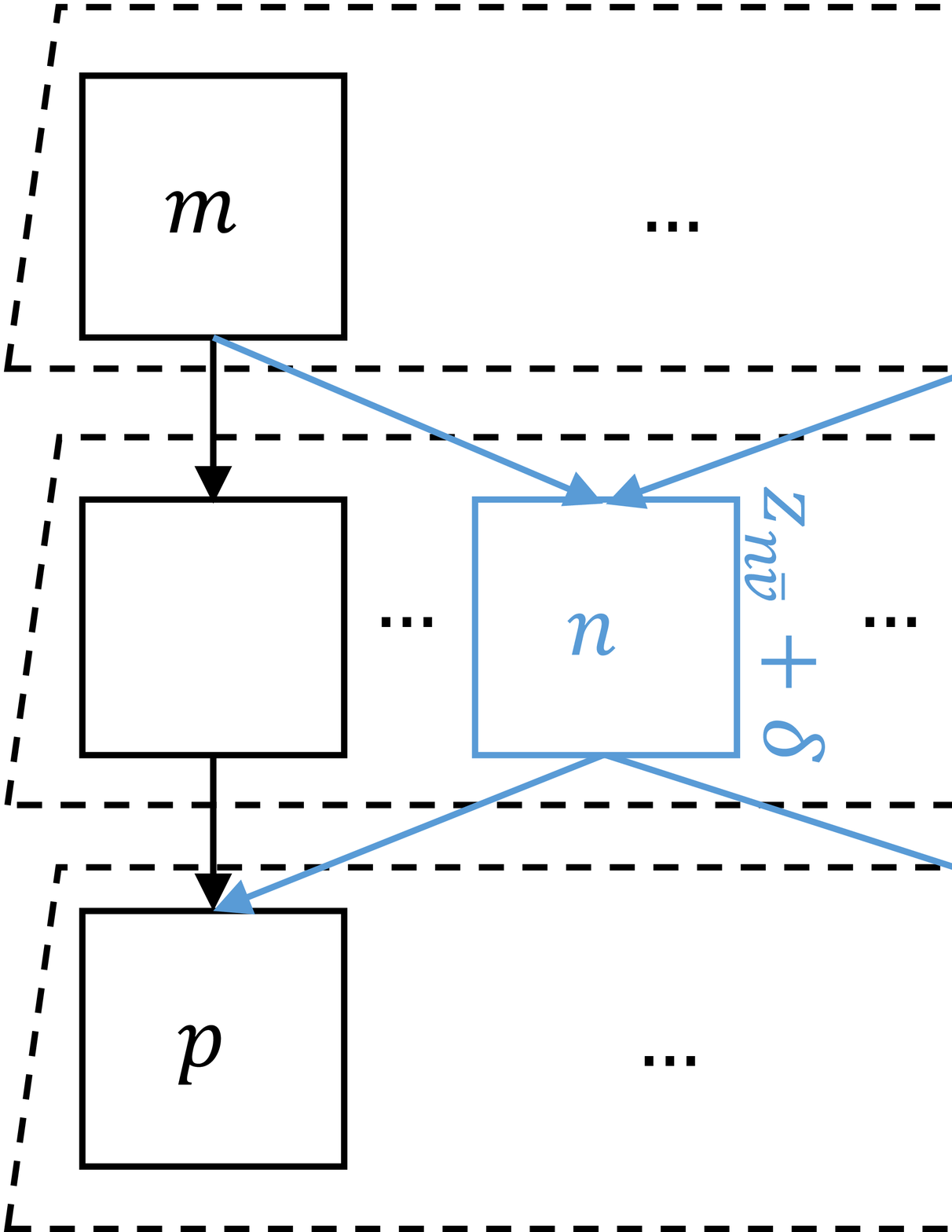}
        \caption{$add(n, L(\overline{v}), \delta)$}
    \label{fig:add}
    \end{subfigure}
    \begin{subfigure}[b]{0.21\textwidth}
        \includegraphics[width=\textwidth]{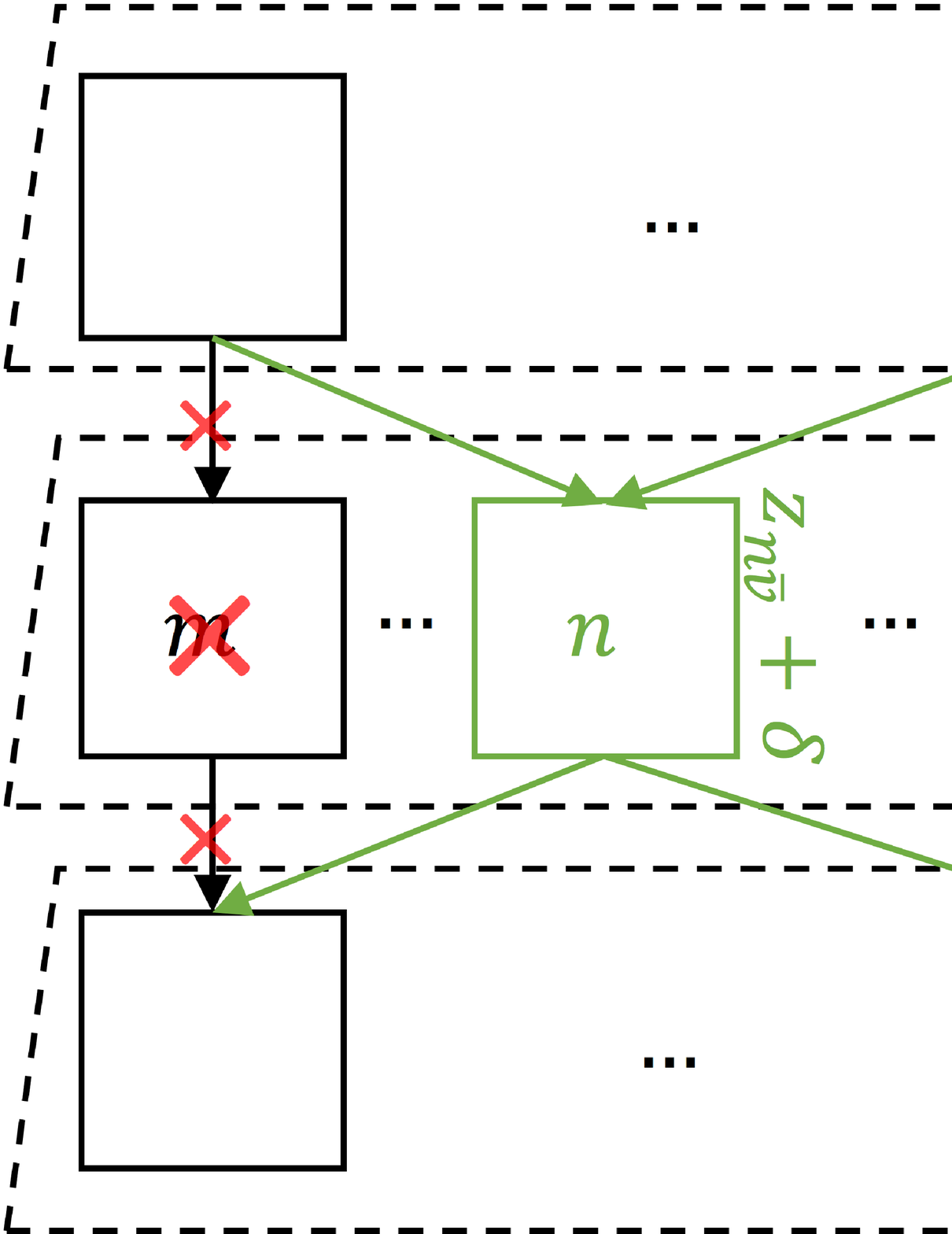}
        \caption{$open(n,\{m\}, L(\overline{v}), \delta)$}
    \label{fig:open}
    \end{subfigure}
    \caption{Actions}
\label{fig:actions}
\vspace{-0.6cm}
\end{figure}

%% file: sections/evaluation.tex
\section{Evaluation}\label{section:evaluation}
\subsection{Experimental Setup}

\subsubsection{Simulated Network}
6-ary Fat-tree \cite{fattree}, a common data-center topology, is used as the simulated network containing 99 nodes (54 hosts and 45 switches) and 162 links providing full bisection bandwidth. Hosts are equipped with 8 core CPU and 1 Gbps network adapter.
The link capacities are 1 Gbps.
The relative importance of allocating 1 Mbps of bandwidth over one link vs. one core CPU is $1\%$.

\subsubsection{VNFs}
We select firewall, IDS, IPsec and WAN-opt. as SFs. \tabl{common-vnfs} reveals the VNFs used in the simulation. Since CPU is the most restricted host resource while dominating the cost, we ignore memory and storage requirements.

\input{tables/common-vnfs}

\subsubsection{Service Chains}
Sources and targets are uniformly distributed in the data-center network. Poisson distribution with the average of 1-chain per 100-seconds is used to simulate the arrival rate. Chains lifetimes follow the exponential distribution with an average of 3 hours.

\subsubsection{Parameters}\label{section:params}
We asses Kariz in respect to \textit{throughput-demand} and \textit{length} of chains. In each experiment, the throughput-demand is fixed to one of \{100,150,200,250,300\} Mbps, and one of the following chains is selected.
\begin{itemize}
    \item Len-1: \{Firewall\},
    \item Len-2: \{Firewall $\rightarrow$ IDS\},
    \item Len-3: \{Firewall $\rightarrow$ IDS $\rightarrow$ IPSec\}, and
    \item Len-4: \{Firewall $\rightarrow$ IDS $\rightarrow$ IPSec $\rightarrow$ WAN-opt.\}
\end{itemize}
Note that Len-$i$ contains all SFs of Len-$i\text{-}1$. We consider Len-1 and Len-2 as homogeneous chains because firewall and IDS VNFs in Len-2 almost demand the same resources for the same throughput. Len-3 and Len-4 are more heterogeneous due to different resource requirements of corresponding VNFs.

\subsubsection{Evaluation Method}
We compare Kariz against the optimal solution implemented using CPLEX. We refer to the optimal solution by \textit{MIP}. The tuning parameter of Kariz is set to $\epsilon = 20$. Thus, an action is performed if it improves the current solution by $5\%$. With fixed parameters, we repeat each experiment $10$ times for different generated $1000$ chains, and report the arithmetic mean. In comparison charts, the ratio of Kariz's to MIP's corresponding value is reported.

\subsection{Acceptance Ratio}
The acceptance ratio results are shown in \fig{acc-ratio}. \fig{kariz-acc-ratio} and \fig{mip-acc-ratio} depicts the acceptance ratio of Kariz and MIP, respectively. The values are the average of acceptance ratios of 10 experiments. As expected, the longer chains with higher throughput-demand have the less chance to be accepted. The low acceptance ratio for Len-4 is due to resource hungriness of these chains, especially for WAN-opt. VNFs.

The range of number of accepted chains by Kariz vs. MIP in \fig{comp-acc-ratio} are as follows: 95-100\% for Len-1, 82-95\% for Len-2, 79-100\% for Len-3, and 89-102\% for Len-4. Note that  higher acceptance ratio for Kariz makes sense. Consider a situation that MIP accepts a hard to deploy chain rejected by Kariz. MIP allocates the resources, not allocated by Kariz. Consequently, this allocation prevents MIP from accepting some of the next chains; despite that, Kariz assigns not-allocated resources to these chains resulting in higher acceptance-ratio.

Considering chain length and throughput-demand impacts in \fig{comp-acc-ratio}, Kariz performs closely to MIP. It might be expected that increasing the length of chain and throughput-demand should deteriorate Kariz's acceptance ratio vs. MIP. However, Kariz has better results for Len-3 and Len-4 than Len-2 and Len-1, especially for 250 Mbps throughput-demand. Recall from \sect{solImprove}, Kariz attempts to improve the solution after deployment of every SF of a chain. Since, Len-4 and Len-3 include all SFs of Len-2 and Len-1 chains (see \sect{params}), the expense of more improvement rounds increases the chance of adjusting the earlier solution. All in all, Kariz has a competitive acceptance ratio within 79-100\% vs. MIP.

\input{charts/acc-ratio.tex}

\subsection{Resource Utilization}
Resource utilization of Kariz is compared with MIP in \fig{res-util-comp}. Bandwidth/CPU utilization for Kariz and MIP are the ratio of allocated bandwidth/CPU resources over aggregated bandwidth/CPU capacities in the network. Regarding VNF resources, the reports are the arithmetic mean of per-SF throughput utilization provided by placed VNF instances.

Bandwidth utilization ratios as depicted in \fig{bandwidth-util-comp} are: 97-101\% for Len-1, 88-106\% for Len-2, 78-111\% for Len-3, and 101-131\% for Len-4. \fig{bandwidth-util-comp} and \fig{comp-acc-ratio} shows that Kariz efficiently utilizes the bandwidth resources for Len-1, Len-2, and Len-3 for various throughput-demands. Regarding Len-4, the efficiency of utilizing bandwidth resources is very close to MIP for throughput-demand of 100 and 200 Mbps. However, the efficiency of bandwidth utilization decreases for other throughput demands. 

The CPU utilization ratios are in the range of 95-100\% for Len-1, 84-95\% for Len-2, 76-100\% for Len-3, and 100-103\%, as observed in \fig{cpu-util-comp}. According to \fig{cpu-util-comp} and \fig{comp-acc-ratio}, Kariz utilizes the CPU resources in an efficient way close to MIP.

Finally, the VNF utilization ratios vs. MIP are shown in \fig{vnf-util-comp}. Following ranges are reported: 100-100\% for Len-1, 99-100\% for Len-2, 101-105\% for Len-3, and 101-111\%. Evidently Kariz utilizes VNF instances very closely to MIP for different lengths and throughput demands. 

\input{charts/util.tex}

\subsection{Operational Costs}
\fig{cost} shows Kariz's costs vs MIP. We collect the Kariz's and MIP's average of per chain costs. The reported values are the ratio of Kariz's and MIP's costs. As shown in \fig{bandwidth-cost} on average, Kariz allocates bandwidth resource vs. MIP in the range of: 101-102\% for Len-1, 105-111\% for Len-2, 101-109\% for Len-3, and 100-140\% for Len-4. Regarding CPU as presented in \fig{host-cost}, on average the same number of CPU cores is allocated for Len-1 and Len-2. For Len-3, 0-3\% less number of CPU cores are allocated. Also, 2-13\% more number of CPU cores are allocated to Len-4 by Kariz. Finally, in respect to total operational cost in \fig{total-cost}, following cost ratios vs MIP are observed: 100-101\% for  Len-1, 103-107\% for Len-2, 100-105\% for Len-3, and 99-125\% for Len-4. Note that it makes sense that Kariz pays 1\% less cost than MIP per Len-4 chains. These solutions accept different number of chains in presence of different available resources. For instance, MIP might accept a chain when the resources are scarce, while Kariz not finding a feasible solution rejects this chain. Consequently, MIP would pay more operational cost in average. In summary, Kariz incurs competitive per-chain cost less than 125\% of MIP.

\input{charts/cost.tex}

%% file: tables/common-vnfs.tex
\begin{table}
    \scriptsize
    \caption{Off-the-shelf VNFs}
    \vspace{-0.2cm}
    \centering
    \begin{tabular}{c|c|c|c}
        Middlebox   & VNF   & Throughput & CPU demand
        \\
        \hline
        \multirow{3}{*}{Firewall\cite{firewalls}}           & Level 1    & 100 Mbps & 1 core\\
                                                            & Level 5    & 200 Mbps & 2 core\\
                                                            & Level 10   & 400 Mbps & 4 core
        \\
        \hline
        \multirow{1}{*}{IDS}                                & Bro \cite{bro}            & 80 Mbps    & 1 core
        \\
        \hline
        \multirow{2}{*}{IPSec\cite{ipsecs}}                 & VSR1001   & 268 Mbps   & 1 core \\
                                                            & VSR1004   & 580 Mbps   & 4 core
        \\
        \hline
        \multirow{2}{*}{WAN-opt.\cite{wan-optimizers}} & CCX770M   & 10 Mbps    & 2 core \\
                                                            & CCX1555M  & 50 Mbps    & 4 core
    \end{tabular}
    \label{table:common-vnfs}
    \vspace{-0.6cm}
\end{table}

%% file: charts/acc-ratio.tex
\begin{figure*}
    \begin{subfigure}[b]{1.0\textwidth}
        \centering
        \includegraphics[width=0.32\textwidth]{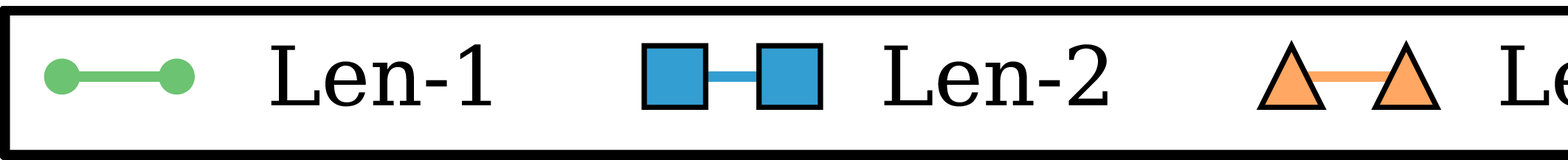}
    \end{subfigure}
    \begin{subfigure}[b]{0.32\textwidth}
        \centering
        \includegraphics[width=\textwidth]{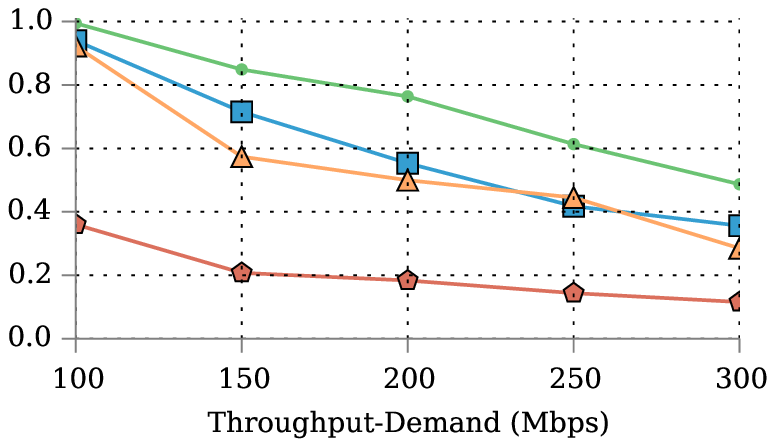}
        \caption{Kariz Acceptance Ratio}
        \label{fig:kariz-acc-ratio}
    \end{subfigure}
    \begin{subfigure}[b]{0.32\textwidth}
        \centering
        \includegraphics[width=\textwidth]{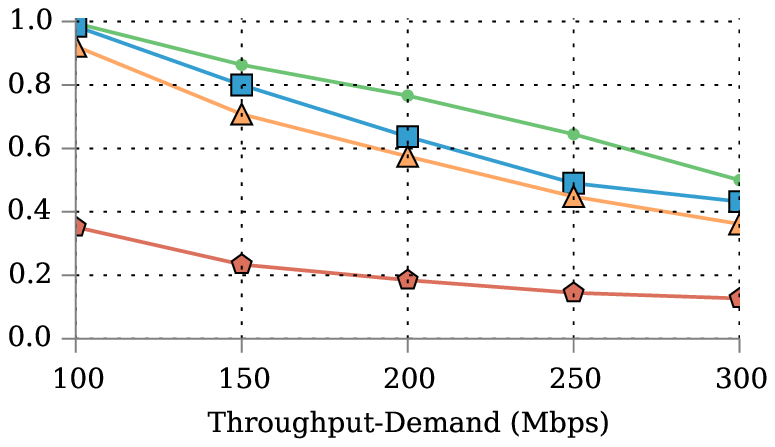}
        \caption{MIP Acceptance Ratio}
        \label{fig:mip-acc-ratio}
    \end{subfigure}
    \begin{subfigure}[b]{0.32\textwidth}
        \centering
        \includegraphics[width=\textwidth]{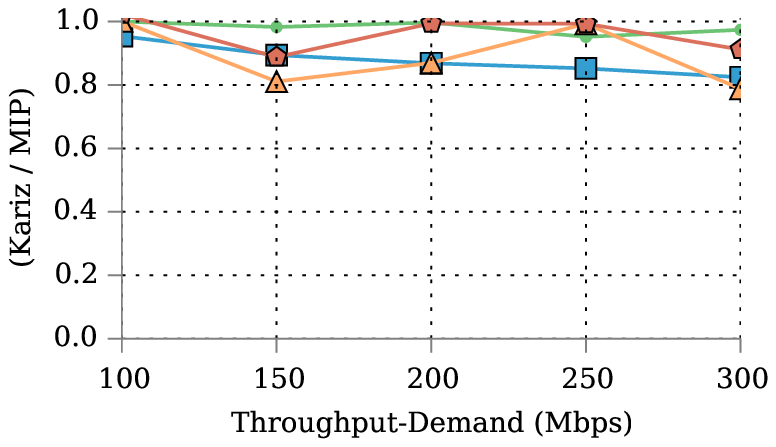}
        \caption{Comparison of Acceptance Ratio}
        \label{fig:comp-acc-ratio}
    \end{subfigure}
    \caption{Acceptance Ratio}
    \label{fig:acc-ratio}
    \vspace{-0.4cm}
\end{figure*}

%% file: charts/util.tex
\begin{figure*}
    \begin{subfigure}[b]{1.0\textwidth}
        \centering
        \includegraphics[width=0.32\textwidth]{charts/legend.eps}
    \end{subfigure}
    \begin{subfigure}[b]{0.32\textwidth}
        \centering
        \includegraphics[width=\textwidth]{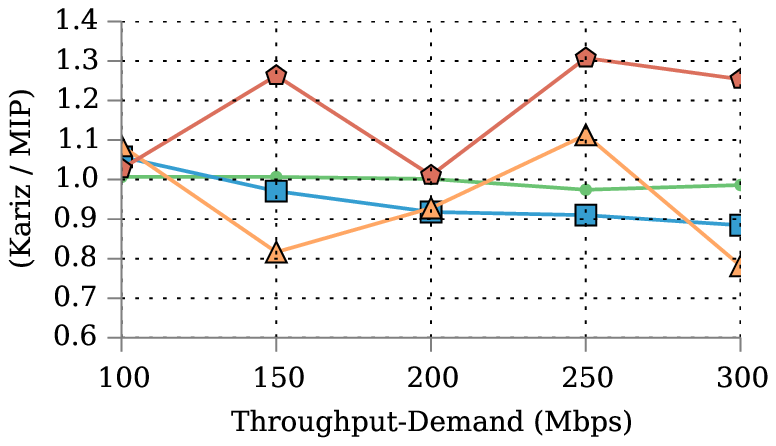}
        \caption{Bandwidth Utilization}
        \label{fig:bandwidth-util-comp}
    \end{subfigure}
    \begin{subfigure}[b]{0.32\textwidth}
        \centering
        \includegraphics[width=\textwidth]{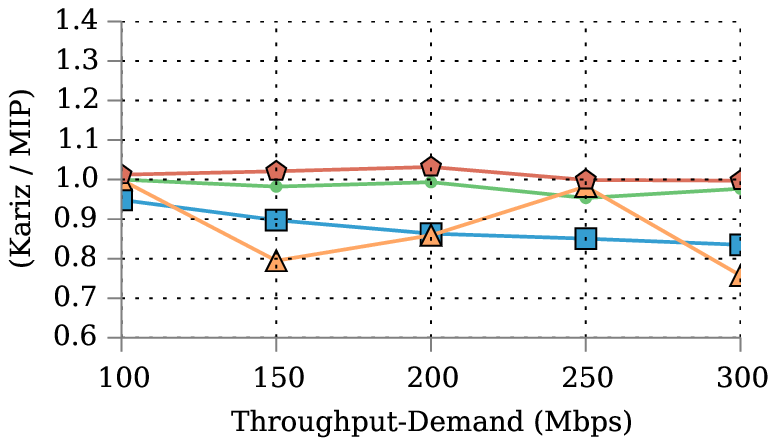}
        \caption{CPU Utilizaiton}
        \label{fig:cpu-util-comp}
    \end{subfigure}
    \begin{subfigure}[b]{0.32\textwidth}
        \centering
        \includegraphics[width=\textwidth]{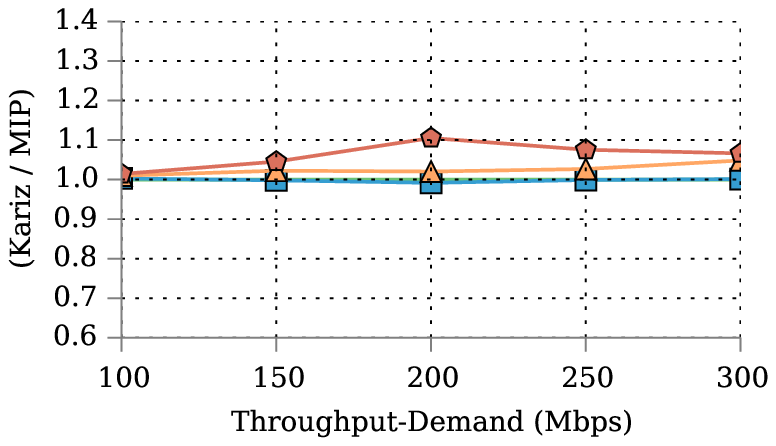}
        \caption{VNF Utilization}
        \label{fig:vnf-util-comp}
    \end{subfigure}
    \caption{Comparison of Resource Utilization}
    \label{fig:res-util-comp}
    \vspace{-0.3cm}
\end{figure*}

%% file: charts/cost.tex
\begin{figure*}
    \begin{subfigure}[b]{1.0\textwidth}
        \centering
        \includegraphics[width=0.32\textwidth]{charts/legend.eps}
    \end{subfigure}
    \begin{subfigure}[b]{0.32\textwidth}
        \centering
        \includegraphics[width=\textwidth]{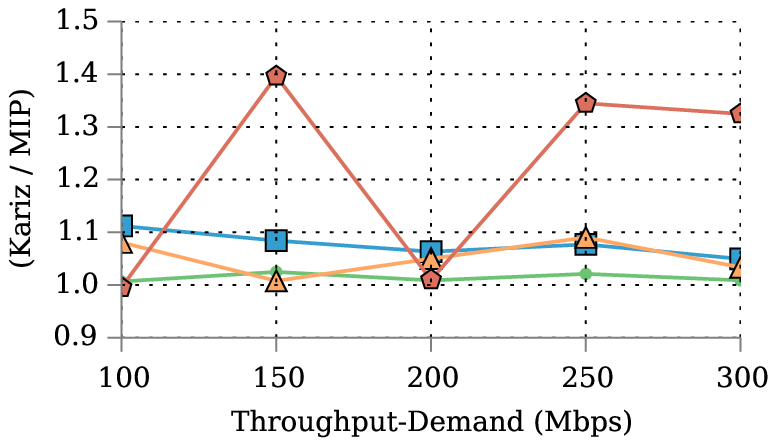}
        \caption{Bandwidth Allocation Cost}
        \label{fig:bandwidth-cost}
    \end{subfigure}
    \begin{subfigure}[b]{0.32\textwidth}
        \centering
        \includegraphics[width=\textwidth]{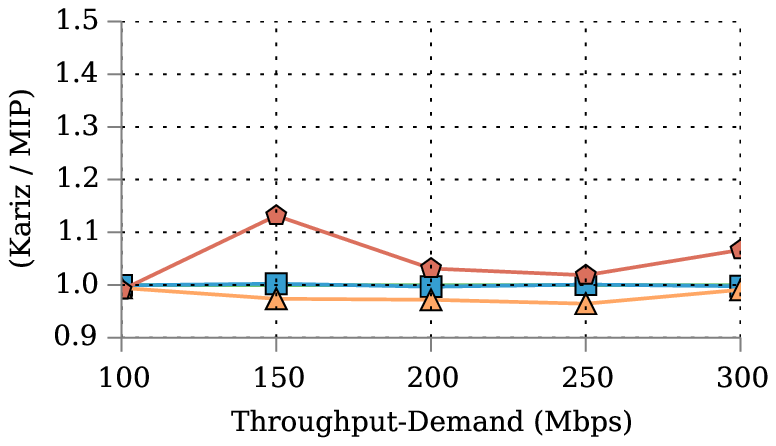}
        \caption{CPU Resource Allocation Cost}
        \label{fig:host-cost}
    \end{subfigure}
    \begin{subfigure}[b]{0.32\textwidth}
        \centering
        \includegraphics[width=\textwidth]{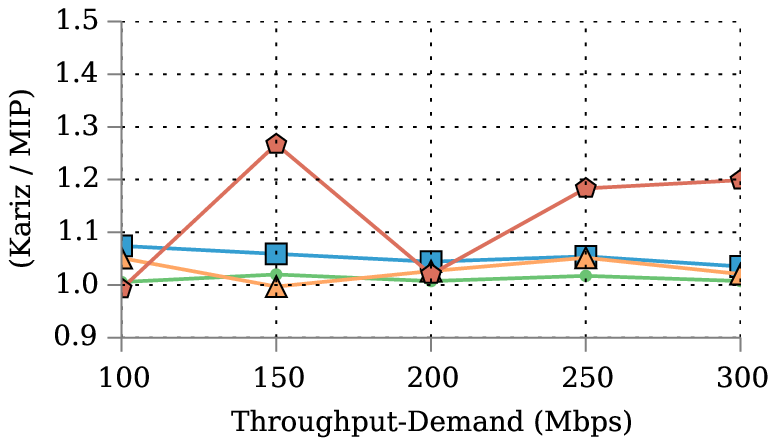}
        \caption{Total Cost}
        \label{fig:total-cost}
    \end{subfigure}
    \caption{Operational Costs}
    \label{fig:cost}
\vspace{-0.6cm}
\end{figure*}

%% file: sections/conclusion.tex
\section{Conclusion}\label{section:conclusion}
Recent optimization models of SFC assume that functionality of a middlebox is provided by a single VNF. This assumption limits SFC to either a single VNF or an individual physical machine. Moreover, heterogeneity of throughput and resource configurations of miscellaneous VNFs makes deployment of a service chain complex. In this paper, we described how we can overcome these limitations. We introduced a mathematical model that enables us to deploy multiple VNF instances to provide the functionality of a middlebox. This eliminates the throughput bound of a chain to a single VNF or a single physical machine. Moreover, this model abstracts heterogeneity of VNFs and allows us to define chains with custom throughput without worrying about individual VNF throughputs. In addition, our Mixed Integer Programming (MIP) model gives the optimal deployment of a chain. For larger scales, we proposed and evaluated a heuristic called Kariz. The experimental results for various chain lengths and throughput demands suggest that Kariz achieves a competitive acceptance ratio of $\sim 80\text{-}100\%$ with an extra cost of less than $25\%$ compared to MIP model. 